\documentclass[11pt]{article}
\usepackage{epsf}
\begin{document}

\def\la{\;\raise0.3ex\hbox{$<$\kern-0.75em\raise-1.1ex\hbox{$\sim$}}\;}
\def\ga{\;\raise0.3ex\hbox{$>$\kern-0.75em\raise-1.1ex\hbox{$\sim$}}\;}
\def\lr{\;\raise0.3ex\hbox{$\rightarrow$\kern-1.0em\raise-1.1ex\hbox{$\leftarrow$}}\;}
\newcommand{\msun}{\mbox{$M_\odot$}}





\markboth{Yakovlev \& Pethick }{Neutron Star Cooling}

\noindent{ \small{To appear in {\it Ann.\ Rev.\ Astron.\ Astrophys.\ } 2004}}

\begin{center}
\LARGE{\bf Neutron Star Cooling}
\end{center}

\noindent{\it D.~G.~Yakovlev $^1$ and C.~J.~Pethick $^2$}

\noindent{$^1$Ioffe Physical Technical Institute,
Politekhnicheskaya 26,
194021 St.-Petersburg, Russia, e-mail: yak@astro.ioffe.ru}

\noindent{$^2$NORDITA, The Nordic Institute for Theoretical Physics, Blegdamsvej 17,
DK--2100 Copenhagen \O, Denmark, e-mail: pethick@nordita.dk}


\begin{abstract}

Observation of  cooling neutron stars
can potentially provide information about
the states of matter at supernuclear densities.
We review physical properties
important for cooling
such as neutrino emission processes and
superfluidity in the stellar interior,
surface envelopes of light elements due
to accretion of matter and strong surface magnetic fields.
The neutrino processes include
the modified Urca process,
and the direct Urca process for nucleons and exotic
states of matter such as a pion condensate,
kaon condensate, or quark matter.
The dependence
of theoretical cooling curves on physical input and observations of thermal
radiation from isolated neutron stars are described.  The comparison of
observation and theory leads to a unified interpretation in terms of three
characteristic types of neutron stars:  high-mass stars which cool primarily
by some version of the direct Urca process; low-mass stars, which cool via
slower processes; and medium-mass stars, which have an intermediate behavior.
The related problem of thermal states of transiently accreting neutron stars
with deep crustal burning of accreted matter is discussed in connection with
observations of soft X-ray transients.

\end{abstract}


\section{INTRODUCTION}
\label{introduc}

Neutron stars are the most compact stars in the Universe.  They have masses
$M \sim 1.4$ \msun\
%
%
and radii $R \sim 10$ km, and they contain matter at
supernuclear densities in their cores.  Our knowledge of neutron star
interiors is still uncertain and, in particular, the composition and equation
of state of matter at supernuclear densities in neutron star cores cannot be
predicted with confidence.  Microscopic calculations are model dependent and
give a range of possible equations of state (e.g., Lattimer \& Prakash 2001;
Haensel 2003), from stiff to soft ones, with different compositions of the
inner cores (nucleons, pion or kaon condensates, hyperons, quarks).

One of the strong incentives for studying the thermal evolution of
neutron stars is the promise that, by confronting observation and
theory, one may learn about
matter in the stellar interior.
The foundation of the theory of neutron star cooling was laid
by Tsuruta \& Cameron (1966). The development of the
theory was reviewed in the 1990s,
e.g., by Pethick (1992), Page (1998a, b), Tsuruta (1998),
and Yakovlev et al.\ (1999). The latter authors presented also
a historical review covering earlier studies. Some recent results have
been summarized by Yakovlev et al.\ (2002b, 2004a).
In this paper we shall discuss the current state of
the cooling theory and compare it with observations
of thermal radiation from isolated neutron stars. We shall
also outline the related problem
of accreting neutron stars and its application
to the quiescent radiation from
soft X-ray transients. We shall describe mainly
results that have been obtained since the middle of 1990s.
The emphasis in this review is on discussing the
dependence of the cooling behavior of neutron stars on physical
properties of the matter and comparing the results with observation, rather
than giving a detailed account of the results of microscopic theory.

\section{OBSERVATIONS}
\label{observ}

Let us begin by describing observations of thermal
radiation from isolated neutron stars --
cooling neutron stars which are not reheated by accretion.
The detection of the thermal radiation
is a complicated problem. Active processes in magnetospheres
of young pulsars (with ages $t \sim 1000$ yrs)
result in strong non-thermal emission
which obscures the thermal radiation. Old pulsars
($t\ga 10^6$) may have hot polar spots due to
the pulsar activity, and the radiation from these
spots
may be stronger than the thermal radiation from the rest of the stellar
surface, which is colder. This complicates the extraction of
the thermal radiation component from the observed spectra.

\begin{figure}[!t]        
\begin{center}
\leavevmode \epsfysize=8cm
 \epsfbox[25 165 550 650]{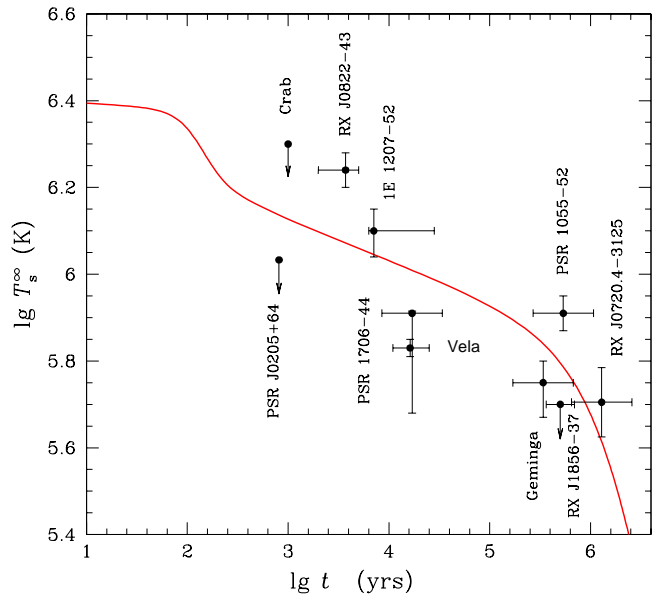}
\end{center}
\caption{Observations of surface temperatures and upper bounds for
several isolated neutron stars. The solid line is
the basic theoretical cooling curve of a nonsuperfluid neutron star with
$M=1.3\,\msun$ (Sect.\ \ref{nosf}).}
\label{fig:observ}
\end{figure}

As a result of these difficulties, thermal radiation
has been reliably detected only from several isolated middle-aged
neutron stars ($t \sim 10^4-10^6$ yrs),
where it appears to be an appreciable fraction of the total radiation.
The main parameters of these stars
(rotation period $P$, distance $D$, and surface magnetic field $B$)
are listed in Table \ref{tab:param};
ages $t$ and effective surface temperatures $T_{\rm s}$ are
presented in Table \ref{tab:observ} and Fig.\ \ref{fig:observ}
and discussed below.
The surface temperatures are
$T_{\rm s} \sim (0.5-1)\times 10^6$ K.
Thus, the thermal radiation is emitted mainly
in soft X-rays, which can be detected by
orbital X-ray observatories.
The first X-ray observatories
of such a type were {\it Einstein} (1978--1981)
and {\it EXOSAT} (1983--1986).
A very important contribution
to observations of cooling neutron stars was made by the {\it ROSAT}
observatory (1990--1998). A new era
began in 1999 with the launching
of {\it Chandra} and {\it XMM-Newton}, new
X-ray observatories of outstanding capability.
The Rayleigh-Jeans tail of the thermal emission from
some isolated neutron stars has been observed in the optical with
ground-based telescopes.

To compare the observations with theory, we need mainly
the neutron star effective temperatures $T_{\rm s}$ and ages $t$.
Because neutron stars are compact,
the effects of general relativity must be taken into account (see, e.g.,
Shapiro
\& Teukolsky 1983, Thorne 1977). To be specific,  we shall denote
the gravitational mass of the star by $M$ and its circumferential
radius by $R$;  $T_{\rm s}$
will denote the effective temperature and
$L_\gamma = 4 \pi R^2 \sigma T_{\rm s}^4$
the thermal
photon luminosity in the local reference frame of the star. Here $\sigma$
is
the Stefan-Boltzmann constant.
The apparent (redshifted) effective temperature $T_{\rm s}^\infty$
and luminosity $L_\gamma^\infty$, as detected by a distant
observer, are
\begin{equation}
    T_{\rm s}^\infty = T_{\rm s}\,\sqrt{1-r_g/R} \quad {\rm and} \quad
    L_\gamma^\infty = L_\gamma \,(1-r_g/R),
\label{apparent}
\end{equation}
where $r_g=2 G M/c^2 \approx 2.95\, M/M_\odot$ km is
the Schwarzschild radius. One often introduces
the apparent radius $R_\infty=R/\sqrt{1-r_g/R}$ which the
observer would see if the telescope could resolve the star.
The surface temperature
distribution of a magnetized star is nonuniform, and for this
case we shall introduce the mean
effective temperature defined by
$\bar{T}_{\rm s}^4=L_\gamma/(4 \pi R^2 \sigma)$, where $L_\gamma$
is the total thermal luminosity.

The values of $T_{\rm s}$ are obtained by fitting the observed
spectra with theoretical models.
The model spectra usually include
thermal and non-thermal components.
The thermal component is described (e.g., Zavlin \& Pavlov 2002)
either by
a black-body spectrum or by a spectrum provided by
a neutron star atmosphere model (with or without a magnetic field).
The atmosphere models studied theoretically in greatest detail are those
composed
of hydrogen. The depth from which a photon emerges from a hydrogen atmosphere
increases noticeably with photon energy
due to the strong energy dependence of radiative opacities.
Thus, more energetic photons emerge from deeper (and hotter)
layers and make
the spectra harder
than the black-body one, for the same $T_{\rm s}$ (i.e., for the same
thermal flux $\sigma T_{\rm s}^4$).
The values of $T_{\rm s}$ inferred from hydrogen
atmosphere models are typically about one half
of
those
inferred using a black-body spectrum.
Iron atmosphere
models give spectra
whose gross behavior is close to the black-body spectrum, but with
spectral
lines in addition. Theoretical atmosphere models are still far from
perfect, especially for cool stars ($T_{\rm s}$ much below
$10^6$ K)
and for stars with strong magnetic fields ($B \ga 10^{12}$ G),
because of the problems
of ionization equilibrium
and spectral opacities
in the cool and/or strongly magnetized atmospheric plasma.

The parameters used for fitting
observed spectra are: $T_{\rm s}$, $R$, the surface
gravity $g$ (and hence, the stellar mass $M$, which is completely
determined by $R$ and $g$), and the surface chemical composition. One can also
add at least two parameters for the non-thermal radiation component
(described commonly by a power-law spectrum and specified by an
intensity and a power-law index). One should, in addition,
specify the distance to the star and
correct the theoretical spectrum for interstellar absorption,
i.e., specify the column density of interstellar hydrogen. The distances
and column densities can be highly uncertain. In such cases,
they too are
treated as free parameters. Thus, the fitting contains
many parameters. Some of them are determined
from the fitting procedure with large errors. To reduce
the errors, one often fixes certain parameters, for instance
$M$ and $R$. In some cases, one needs to introduce
a second thermal (black-body) radiation component
characterized by its own
effective surface temperature $T_{\rm s1}$ and radius $R_1$.
One usually finds $T_{\rm s1}> T_{\rm s}$ and $R_1 < R$, in which case
the second component has a natural interpretation as thermal emission from
a hot spot on the stellar surface associated with the pulsar activity
and not related to the thermal radiation emerging from
the interior of the star. Stellar ages may also be rather uncertain.
All in all, the observational data in Tables \ref{tab:param}
and \ref{tab:observ}
represent the current state of the art. It has been a
challenge to obtain them, but they are still not very precise.

\renewcommand{\arraystretch}{1.2}
\begin{table*}[!t]   
\caption[]{Isolated
neutron stars which show thermal surface emission}
\label{tab:param}
\begin{center}
\begin{tabular}{|| l | r | c | l | l ||}
\hline
\hline
            Source & $P$ [ms] & $D$ [kpc] &  $B$[10$^{12}$ G] & Comments   \\
\hline
\hline
PSR J0205+6449     & 65      & $\sim 3.2$     & 3.6     &
PSR in SN1181     \\
Crab               &  33     & $\sim 2.5$     & 3.8     &
PSR in SN1054      \\
RX J0822--4300     &  75     & 1.9--2.5 & 6.8     &
CCO in Puppis A       \\
1E 1207.4--5209        & 424     & $2.1^{+1.8}_{-0.8}$ & 4  &
CCO in G296.5+10.0 \\
Vela               &  89     & $0.293^{+0.019}_{-0.017}$$^{~a)}$ & 3.4 &
PSR \\
PSR B1706--44      &  102    & $\sim 2.3$ &          3      &
PSR \\
Geminga            & 237     & $0.159^{+0.059}_{-0.034}$$^{~a)}$ & 1.6 &
Musketeer \\
RX~J1856.4--3754     &         & $117\pm12$$^{~a)}$     &    &
Dim object\\
PSR~B1055--52      &  197    & $\sim 0.9$     & 1.1 &
Musketeer  \\
RX J0720.4--3125   & 8391    & $\sim 0.2$     & 9.3 &
Dim object\\
\hline
\end{tabular}
\end{center}
\small{
$^{a)}$ parallax measured\\
}
\end{table*}
\renewcommand{\arraystretch}{1.0}

\renewcommand{\arraystretch}{1.2}
\begin{table*}[!t]   
\caption[]{Observational limits on surface temperatures of isolated
neutron stars}
\label{tab:observ}
\begin{center}
\begin{tabular}{|| l | c | c | c | l ||}
\hline
\hline
Source & $t$ [kyr] & $T_{\rm s}^\infty$ [MK] &  Confid.\  & References   \\
\hline
\hline
PSR J0205+6449     & 0.82    & $<$1.1$^{~b)}$  &      & Slane et al.\ (2002)    \\
Crab               &    1    & $<$2.0$^{~b)}$   &  99.7\%    & Weisskopf et al.\
(2004)  \\
RX J0822--4300     & 2--5    & 1.6--1.9$^{~a)}$ & 90\% & Zavlin et al.\ (1999)   \\
1E 1207.4--5209        & $\ga$7 & 1.1--1.5$^{~a)}$ & 90\% & Zavlin et al.\ (1998) \\
Vela               & 11--25  & 0.65--0.71$^{~a)}$ & 68\% & Pavlov et al.\ (2001)\\
PSR B1706--44       & $\sim$17 & 0.82$^{+0.01}_{-0.34}$$^{~a)}$ & 68\% &
McGowan et al. (2004) \\
Geminga            & $\sim$340 & $0.56^{+0.07}_{-0.09}$$^{~b)}$ & 90\% &
Halpern \& Wang (1997) \\
RX~J1856.4--3754     & $\sim$500 & $<$0.5  & -- & Pavlov \& Zavlin (2003) \\
PSR~B1055--52       & $\sim$530 & $\sim$0.7$^{~b)}$ & -- &
Pavlov (2003)  \\
RX J0720.4--3125   & $\sim 1300$ & $\sim 0.5$$^{~a)}$  & -- &
Motch et al.\ (2003) \\
\hline
\end{tabular}
\end{center}
\small{
$^{a)}$ Inferred using a hydrogen atmosphere model\\
$^{b)}$ Inferred using the black-body spectrum\\
}
\end{table*}
\renewcommand{\arraystretch}{1.0}

Isolated (cooling) neutron stars do not constitute a homogeneous class of
objects. The two youngest objects, PSR J0205+6449 and the famous
Crab pulsar (PSR B0531+21), are radio pulsars in
historical supernova remnants.
The next two objects, RX J0822--4300 and
1E 1207.4--5209 (=J1210--5226), are radio-quiet
neutron stars in the supernova remnants Puppis A and
G296.5+10.0 (=PKS 1209--51/52). They belong
to the class of radio silent compact central objects
(CCOs) in supernova remnants. CCOs have recently been reviewed
by Pavlov et al.\ (2002b) and Pavlov \& Zavlin (2003).
1E 1207.4--5209 is the first isolated neutron star
found to exhibit pronounced spectral features
(spectral lines) in its radiation spectrum
(Sanwal et al.\ 2002),
although the interpretation of these features seems ambiguous.
The next member of the list, the Vela pulsar (PSR B0833--45), is almost as
famous as the Crab pulsar. Among other objects, there are 
three
neutron stars
-- PSR B1706--44, 
the Geminga pulsar (PSR B0633+1748), and PSR B1055--52 --
which have been observed as radio pulsars.
The properties of the radiation of 
two
of them, 
Geminga and PSR B1055--52,
and of another source, PSR B0656+14  (not
included in our analysis as explained below),
are so similar that
J.\ Tr\"umper has
dubbed them the  {\it three musketeers}.
The other two sources in Tables
\ref{tab:param} and \ref{tab:observ}, 
RX J1856.4--3754 and RX J0720.4--3125,
are radio-silent, nearby, {\it dim}, isolated neutron stars observed in the
optical band, UV and X-rays. The properties of dim objects have been reviewed
by Pavlov \& Zavlin (2003).

The {\it rotation periods $P$} have been
measured (Table \ref{tab:param}) for all the sources but
RX J1856.4--3754 in various spectral bands (particularly,
in the radio for radio pulsars and in X-rays); the measured
(or estimated) time
derivatives $\dot{P}$
show a familiar neutron star spindown.
The {\it magnetic fields} $B$ on neutron star surfaces have been determined
in the standard way from $P$ and $\dot{P}$.
Numerous attempts to detect the rotation
 of RX J1856.4--3754 by observing periodic variations of the radiation
have failed.

The {\it distances} $D$ to the neutron stars in supernova remnants
(PSR J0205+6449, Crab, RX J0822--4300, 1E 1207.4--5209)
are the estimated distances to the remnants.
The distances to the Vela pulsar (Caraveo et al.\ 2001,
Dodson et al.\ 2003), 
Geminga (Caraveo et al.\ 1996), and RX J1856.4--3754 (Walter \&
Lattimer 2002) are reliably known from parallax measurements.
In all other cases except RX J0720.4--3125, $D$
has been determined
by standard methods of radio astronomy and, consequently,
the uncertainty may be large.
We take the distance to
RX J0720.4--3125 from Motch et al.\ (2003) as estimated
by fitting the observed spectrum with a neutron star atmosphere model
(see below).

The {\it ages} (Table \ref{tab:observ}) of PSR J0205+6449 and the Crab pulsar
are the well known ages of the historical supernovae.
For RX J0822--43,
we take the estimated age $t=2-5$ kyr of the host
supernova remnant (as can be deduced, e.g.,
from a discussion in Arendt et al.\ 1991), with a central value of
$t=3.7$ kyr (Winkler et al.\ 1988).
For 1E 1207.4--5209 (as in Yakovlev et al.\ 2004a), we adopt 
the age range from the
standard age of the associated supernova remnant $\sim 7$ kyr
to a value four times longer. For Vela, we take the
age interval from the standard spindown age $t_{\rm c}$ to
the age estimated by Lyne et al.\ (1996) taking into account
variations of $\dot{P}$ due to Vela's glitches.
The age of RX J1856.4--3754 has been revised recently
by Walter \& Lattimer (2002)
from the analysis of star's proper motion;
the error bar is chosen to
distinguish clearly the revised value from the previous one.
The ages of other neutron stars are the characteristic pulsar spindown
ages $t_{\rm c}$ assuming an uncertainty of a factor of 2.
In particular, we take $\dot{P}$ for RX J0720.4--3125
from the recent measurements of Kaplan et al.\ (2002).

Finally, {\it effective surface temperatures} $T_{\rm s}^\infty$
in Table \ref{tab:observ} are especially delicate.
Unfortunately, no thermal radiation has been detected
from the youngest objects,
PSR J0205+6449 and the Crab pulsar, since their thermal emission
is obscured by
a strong non-thermal one. Nevertheless, upper
limits on $T_{\rm s}^\infty$ for these objects have been obtained by
Slane et al.\ (2002) and Weisskopf et al.\ (2004).

The surface
temperatures of RX J0822--4300 and 1E 1207.4--5209 in Table
\ref{tab:observ}
are taken from Zavlin et al.\ (1999, 1998).
They have been obtained with the aid of hydrogen atmosphere models.
Such results are more consistent with other data and
theoretical predictions (with estimated
distances to the sources, interstellar hydrogen column
densities, theoretical neutron star radii) than are results based
on the black-body spectrum.
Recently the values of $T_{\rm s}^\infty$ for these
and some other isolated neutron stars have been revised, for instance, by
Pavlov et al.\ (2002b) and Pavlov \& Zavlin (2003).
The revised values are in line with the previous ones
but they are presented by the authors without error bars. Accordingly,
we prefer to use the previous values given
with the estimated uncertainties.

The surface temperatures of the Vela
pulsar and PSR B1706--44 are taken from Pavlov et al.\
(2001) and McGowan et al.\ (2004), respectively.
Again they have been inferred using hydrogen atmosphere models.
The surface temperature of PSR B1706--44 is much less
certain because the source is more distant,
but the value may be improved in the future.

The values of $T_{\rm s}^\infty$ for two musketeers -- Geminga and
PSR B1055--52 -- are taken, respectively, from Halpern \& Wang (1997)
and Pavlov (2003), who obtained them with the black-body
model, which seems more appropriate for these
sources. The values of $T_{\rm s}^\infty$ for PSR B1055--52
have been reported without error bars, but in
Fig.\ \ref{fig:observ} we include, somewhat arbitrarily, 10\% uncertainties
in $T_{\rm s}^\infty$. Let us remark also on the
third musketeer, PSR B0656+14. 
Recent parallax measurements for this pulsar
by Brisken et al.\ (2003) have yielded the distance
$D \approx 290$ pc, noticeably smaller than accepted before.
Multiwavelength observations of the source have been
reported most recently, e.g., by Zavlin \& Palvov (2004).
It is difficult to properly extract thermal luminosity
emitted from the entire neutron star surface using
either black-body or hydrogen atmosphere spectra. 
Thus, we do not include PSR B0656+14 in our analysis.

The surface temperature of RX J1856.4--3754 is still poorly
determined. The thermal X-ray component is well fitted by
the black-body spectrum, but its extrapolation to the optical band
(the Rayleigh-Jeans tail) strongly underestimates the observed optical flux.
It is possible that the star has a hot spot (or spots)
whose radiation contaminates the emission from
the rest of the surface,
which is cooler. Generally, the temperature distribution
over the surface may be nonuniform and the
true average surface temperature is poorly determined. The nondetection
of pulsed emission may be due to
unfavorable orientations of radiation beams and/or
due to
strong gravitational lensing, which may smear out the
anisotropy of the radiation. For definiteness, we have taken
a model nonuniform temperature distribution
suggested by Pavlov \& Zavlin (2003) to fit the observations.
We have calculated
the appropriate average surface temperature,
determined by the total photon luminosity
(see above),
to be
$T_{\rm s}^\infty \approx 0.5$ MK. We propose to treat
this value as an upper limit on $T_{\rm s}^\infty$.
Such a limit does not contradict other estimates
of $T_{\rm s}^\infty$ for this source
(e.g., Pons et al.\ 2002, Braje \& Romani 2002, Burwitz et al.\ 2003).

The surface temperature of the last source,
RX J0720.4--3125, is taken from Motch et al.\ (2003),
who fitted the observed spectrum with a model
of a hydrogen atmosphere of finite depth.
The authors do not report any error bars, but
we have included 20\% uncertainties which we think
are appropriate for this case.

The observational data are displayed in Fig.\ \ref{fig:observ}: they are
a scatter of {\it sparse observational limits}. In the next
sections we shall outline the theory of neutron star cooling  and
discuss the main issue: {\it what we can learn about
the internal structure of neutron stars by confronting
theory and observation}.

\section{THEORETICAL OVERVIEW}
\label{theory}

\subsection{Internal Structure}
\label{internal structure}

Current theories of the internal structure of neutron stars are described,
e.g., by Glendenning
(1996), Weber (1999), Lattimer \& Prakash (2001), and Haensel (2003).  A
neutron star has a thin atmosphere and four internal regions, which we shall
refer to as the {\it outer crust}, the {\it inner crust}, the {\it outer
core}, and the {\it inner core}.

In the {\it outer crust} (the outer envelope),
 matter consists of ions (atomic nuclei) and electrons.
The electrons constitute a strongly degenerate,
almost ideal gas, which is relativistic at densities of order 10$^6$ g
cm$^{-3}$ and above.
The ions form a strongly coupled
Coulomb system, which is solid in most of the envelope, but which
is
liquid at the lowest densities.
The electron Fermi energy grows with increasing $\rho$ and, as a consequence,
nuclei tend to become richer in neutrons since it is energetically
favorable to convert electrons and protons into neutrons (and neutrinos) by
electron capture processes.
The nuclei can also undergo other
transformations, such as
pycnonuclear reactions,
absorption and emission
of neutrons.  At the base of the outer crust,
neutrons begin to drip out of nuclei, thereby
producing a neutron gas between nuclei.  This occurs at a density
$\rho = \rho_{\rm ND} \approx 4 \times 10^{11}$ g~cm$^{-3}$.
The thickness of the outer crust is a few hundred meters.

In the {\it inner crust} (the inner envelope), matter consists of electrons,
free neutrons and neutron-rich atomic nuclei (Negele \& Vautherin 1973,
Pethick \& Ravenhall 1995, Haensel 2003).  The fraction of free neutrons
increases with increasing $\rho$, and at the bottom of the crust (in the
density range from $\approx 10^{14}$ to about $\rho_0/2$), where nuclei occupy
a significant fraction of space, nuclei may be far from spherical (Lorenz et
al.\ 1993, Pethick \& Ravenhall 1995), but the detailed structure depends on
the nuclear model used (Oyamatsu 1993).  Here, $\rho_0 \approx 2.8 \times
10^{14}$ g cm$^{-3}$ is the density of nuclear matter at saturation.  Nuclei
disappear at a density $\sim 0.5\rho_0$, and matter then becomes a uniform
fluid of
neutrons,
protons and electrons.  The thickness of the inner crust is
typically about one kilometer.  The equation of state throughout the crust has
been calculated with reasonable accuracy, and is sufficiently well understood
for the purpose of building neutron star models.

Below the inner crust lies the stellar {\it core}.  At the lowest densities
matter consists of neutrons with an admixture (several per cent by particle
number) of protons, electrons, and possibly muons (the so called $npe$ and
$npe\mu$ compositions).  All constituents of dense matter are strongly
degenerate.  The neutrons and protons, which interact via nuclear forces,
constitute a strongly non-ideal liquid.  For densities up to about $ 2 \rho_0$
the equation of state and composition of matter are reasonably well
constrained by nuclear physics data and theory, while at higher densities they
are much
less certain.
We shall divide the core into an {\it outer core} at
densities less than $\sim 2 \rho_0$ and an {\it inner core} at higher
densities.  The distinction between the inner and outer cores
reflects
our ignorance concerning the state of matter at high density, and does
not imply that matter is physically different in the two parts of the core,
although it could be. In
massive stars the radius of the
inner core may reach several kilometers,
and its central density may be as high as $(10-15) \rho_0$, while in low-mass
stars the inner core may be absent since the outer core extends to
the center of the star.

There are a number of possibilities for the composition of dense matter:

{\it (1)}
{\it Nucleon matter.}
The constituents of matter are basically the
same as in the outer core.

{\it (2)}
{\it Hyperonic matter.}
With increasing density, the neutron and
electron chemical potentials also increase,
exceeding possibly thresholds for creating
heavier particles such as
$\Sigma^-$ and $\Lambda$ hyperons.

{\it (3)}
{\it Pion condensate.}
In dense matter, the energy of a $\pi$
meson
(pion) is modified by interparticle interactions and, if it becomes
sufficiently low, a Bose-Einstein condensation of
pion-like excitations can appear (Migdal 1971, Sawyer 1972, Scalapino 
1972, Baym \& Campbell 1978).

{\it (4)}
{\it Kaon condensate.}
The energy of a $K$ meson (kaon) will
likewise be affected by interactions, and therefore another possibility is a
Bose-Einstein condensation of kaon-like excitations (Kaplan \& Nelson 1986,
Brown 1995) which, like real kaons, possess strangeness.

{\it (5)}{\it Quark matter.} At high densities, it is predicted that
nucleons will merge to make a fluid composed of light $u$ and $d$ quarks and
strange $s$ quarks, and a small admixture of electrons, so-called quark
matter (see Weber 1999, for a review). The appropriate degrees of freedom
of such matter are then quarks and gluons, rather than nucleons and mesons.
Because of the presence of $s$ quarks, the matter is sometimes referred to
as strange quark matter.

The reason that the state of matter at high density plays such an important
role in the cooling of neutron stars is that during its early life, a neutron
star
cools primarily by neutrino emission.  Neutrino emission depends sensitively
on the nature of the low-lying excitations in matter, and these are different
for the different types of matter.  Thus, observations of neutron star cooling
may provide a way of discriminating between the various possible states of
matter at high densities.  The nature of matter also affects neutron star
models, since the appearance of a new phase tends to soften the equation of
state, and thereby alter the structure of a
star.

Nuclear matter
without and
with hyperons have been studied
experimentally in ordinary nuclei and hypernuclei.  Neither boson
condensation nor the
deconfinement of nuclear matter have yet been discovered in nuclei in the
laboratory so far.
Thus, the last three models ({\it 3}) -- ({\it 5}) are often referred to as
{\it exotic} models of dense matter.  One cannot exclude the existence of
mixed phases.

The numerous theoretical equations of state may be subdivided into three
classes, {\it soft},
{\it moderate}
and {\it stiff} ones, with respect to the {\it compressibility} of matter.
Employing different equations of state, one obtains different stellar models
and, in particular, different maximum masses of a stable neutron star, from
$M_{\rm max} \sim 1.4 \, \msun$ for the softest equations of state to $M_{\rm
max}
\sim
(2-2.5) \, \msun$ for the stiffest ones.
The appearance of a new phase tends to soften the equation of state and,
consequently,
the compressibility and composition are correlated. Generally
speaking, very stiff equations of state are found only for nucleon matter.

Let us also mention a special hypothetical class
of compact stars, called {\it strange} stars,
which consist almost entirely of strange quark matter.
According to some models, this matter extends to the very surface;
such stars are referred to as {\it bare strange stars}.
According to other models, strange stars may have a normal crust
extending from the surface to the neutron-drip
density. Strange stars are discussed in detail by
Weber (1999).

\subsection{Superfluidity}
\label{superfluidity}

Superfluidity of nucleons in atomic nuclei is well established. 
Migdal (1959) predicted that neutrons in neutron stars would be superfluid.
Other baryons 
could also be superfluid there.
The superfluidity
is produced by pairing (so-called Cooper pairing) of baryons due to the
attractive component of their interaction, and it is present only
when the temperature $T$ of the matter falls below a critical temperature
$T_{\rm c}(\rho)$, specific for a particular baryon species.
An important microscopic effect is that the onset of superfluidity leads to
the appearance of a gap
$\Delta$
in the spectrum of elementary excitations near the Fermi surface.  The pairing
phenomenon is largely confined to states in the vicinity of the Fermi surface
and, consequently, it has almost no effect on the equation of state, and hence
on neutron star masses and radii.  Superfluidity of charged
particles (for instance, protons) implies that matter is superconducting.

Theory predicts
that the neutrons between nuclei in the inner
crust (and nucleons in nuclei) will pair in the
spin {\it singlet state}
($^1$S$_0$) with zero orbital angular momentum, just as electrons do in
conventional metallic superconductors.
However, this neutron pairing
is expected to
disappear
in the stellar core because the neutron-neutron
interaction
in the singlet state becomes repulsive
with increasing density  (Wolf 1966). 
Nevertheless, a weaker interaction
in the spin triplet state ($^3$P$_2$) with unit orbital angular momentum
may be still attractive in the core
producing the {\it triplet-state pairing}
of neutrons with an anisotropic gap
(Hoffberg et al.\ 1970). 
(This state is a close relative of
those for the superfluid phases of liquid helium 3.) 
Purely $^3$P$_2$ pairing could be a simplification, because
$^3$P$_2$ channel can be superimposed with other ones, particularly,
with $^3$F$_2$.
Owing to a much lower
number density
of protons in the core, their pairing occurs usually in the singlet state.
Hyperons
(Balberg \& Barnea 1998)  
and quarks
(Bailin \& Love 1984) 
could also be superfluid.
Pion and kaon condensations affect superfluidity of nucleons
(e.g., Takatsuka \& Tamagaki 1995, 1997).

Critical temperatures $T_{\rm c}$ of various
particle species have been calculated by many
authors as reviewed by
Lombardo \& Schulze (2001)
(more references can be found in
Yakovlev et al.\ 1999).
The results are extremely sensitive to the strong interaction
models and many-body theories employed.
In all the cases mentioned above, microscopic theories
give density dependent critical
temperatures $T_{\rm c} \la 10^{10}$~K.
All
types of pairing are expected to
disappear at
densities well above nuclear density,
where the attractive part
of
the strong interaction becomes ineffective.

In addition,
Alford et al.\ (1998) 
proposed a new type of quark
superfluidity. It consists in pairing of unlike quarks
($ud$, $us$, $ds$)
in states which possess not only flavor but also color, in contrast to the
quark pairing described above (in which a pair carries neither color nor
flavor).
This is
referred to as
{\it color superconductivity}.
For a typical quark Fermi energy
$\sim 500$ MeV, one may expect critical temperatures
$T_{\rm c} \sim 50$ MeV $\sim 5 \times 10^{11}$ K.
This topic is an active area of research, and properties of stars composed of
such quark matter are under investigation (see, e.g., Alford 2004 and
references therein).

Because of the energy gap, superfluidity
has marked effects on
the heat capacity and the neutrino emissivity of dense
matter.  It induces also a number of macroscopic quantum phenomena (quantized
neutron vortices in rotating neutron stars and quantized magnetic flux tubes
in magnetized
neutron-star cores, --
Baym et al.\ 1969)
but, since they are less central to the cooling of neutron stars,
we do not discuss them in this review.

\subsection{Neutrino Emission Processes}
\label{neutrinos}

Neutrino emission is generated in numerous reactions
in the interiors of neutron stars, as reviewed, for instance, by
Pethick (1992) and Yakovlev et al.\ (2001a).
Neutrinos carry away energy
and provide efficient cooling of warm neutron stars
with internal temperature $T \ga (10^6-10^7)$ K.
Neutrino reactions in the stellar crust are
summarized by Yakovlev et al.\ (2001a).
The most powerful neutrino emission is produced in
the stellar core. Typical neutrino energies
in nonsuperfluid stars are $\sim k_{\rm B}T$, where $k_{\rm B}$ is
Boltzmann's constant.

The neutrino mechanisms in the core can be subdivided
into {\it slow} and {\it fast} ones.
In nonsuperfluid dense matter,
the emissivities of slow and fast neutrino processes
can be written as
\begin{equation}
   Q_{\rm slow}= Q_{\rm s} T_9^8,\qquad
   Q_{\rm fast}= Q_{\rm f} T_9^6,
\label{Qnu}
\end{equation}
where $T_9=T/(10^9~{\rm K})$, while $Q_{\rm s}$ and $Q_{\rm f}$ are
slowly varying functions of $\rho$ (whose estimates are
presented in Tables \ref{tab:fast} and \ref{tab:slow}
taken from Yakovlev \& Haensel 2003 with kind permission of the
authors).

\newcommand{\rrr}{\rule{0cm}{0.3cm}}
\begin{table*}[!t]
\caption{Processes that provide fast
neutrino emission
in nucleon matter and three models of exotic matter}
\begin{center}
  \begin{tabular}{|lll|}
  \hline
  Model              & Process             &
        $Q_{\rm f}$, erg cm$^{-3 \rrr}$ s$^{-1}$ \\
  \hline
  Nucleon matter &
  ${
  n \to p e \bar{\nu} \quad
   p e \to n \nu }$ & $\quad
  10^{26 \rrr}-3 \times 10^{27}$  \\
  Pion condensate &
  ${
  \widetilde{N} \to \widetilde{N}
  e \bar{\nu} \quad
   \widetilde{N} e \to \widetilde{N}
   {\nu} } $ & $ \quad
  10^{23 \rrr}-10^{26}$  \\
   Kaon condensate &
  ${
  \widetilde{B} \to \widetilde{B}
  e \bar{\nu} \quad
   \widetilde{B} e \to \widetilde{B}
   {\nu} } $ & $ \quad
   10^{23 \rrr}-10^{24}$ \\
   Quark matter &
   ${  
   d \to u e \bar{\nu} \quad  u e \to d \nu } $ & $  \quad
   10^{23 \rrr}-10^{24}$ \\
   \hline
\end{tabular}
\label{tab:fast}
\end{center}
\end{table*}

{\it Fast} neutrino reactions (Table \ref{tab:fast})
have density thresholds
and
they are not expected to
occur outside the inner core.  
The most powerful neutrino emission is
provided by {\it direct Urca}
\footnote{The name {\it Urca} was suggested by Gamow \& Schoenberg
(1941), see
Pethick (1992) and Yakovlev et al.\ (1999) for more details.}
processes in
nucleon or nucleon/hyperon matter (Lattimer et al.\ 1991,
Prakash et al.\ 1992).
An example of a direct Urca process is given in
the first line of Table \ref{tab:fast}. It consists of a pair of reactions,
the beta decay of a neutron  and electron capture on a proton, whose net
effect is the emission of a neutrino--antineutrino pair, the
composition of matter remaining unchanged. The
process can occur only if the proton concentration is sufficiently high.  The
reason for this is that, in degenerate matter, only particles with energies
within $\sim k_{\rm B}T$ of the Fermi surface can participate in reactions,
since other processes are blocked by the Pauli exclusion principle. If the
proton and electron Fermi momenta are too small compared with the neutron
Fermi momenta, the process is forbidden because it is impossible to satisfy
conservation of momentum.
Under typical
conditions one finds that the ratio of
the number density of protons to that of nucleons must exceed about 0.1 for
the process to be allowed.  Proton fractions in the outer core are
estimated to be lower than this, but in the inner core they could be high
enough for the process to
occur.
Other direct Urca
processes may involve muons instead of electrons
and/or hyperons instead of nucleons. In particular,
a concentration of
$\Lambda$ hyperons of order $10^{-3}$ can lead to rapid neutrino emission.

Exotic phases of matter
in the inner cores of massive
stars would lead to fast neutrino emission
by direct-Urca-like processes (see
Pethick 1992, 
for details). Such processes are also
efficient, but somewhat weaker than the nucleon direct Urca one.
The leading processes of such a type in pion-condensed,
kaon-condensed, and quark matter are collected in Table \ref{tab:fast}.
The symbol $\widetilde{N}$ denotes a
nucleon quasiparticle which, in the pion-condensed
phase, is a coherent superposition of a neutron and a proton.  In the
kaon-condensed
state, there are two types of baryonic quasiparticles,
neutron-like ones which are
coherent superpositions of
neutrons and $\Sigma^-$
hyperons,
and proton-like
ones which are coherent superpositions of protons, and  $\Sigma^0$ and
$\Lambda$ hyperons
(both types are denoted collectively by $\widetilde{B}$).

\begin{table*}[!t]
\caption{Main processes of slow neutrino emission
in nucleon matter: modified Urca and
bremsstrahlung}
\begin{center}
  \begin{tabular}{|lll|}
  \hline
  Process   &    &  $Q_{\rm s}$, erg cm$^{-3 \rrr}$ s$^{-1}$ \\
  \hline
  Modified Urca &
  ${
  nN \to pN e \bar{\nu} \quad
   pN e \to nN \nu } $ &
  $\quad 10^{20 \rrr}-3 \times 10^{21}$  \\
  Bremsstrahlung &
  ${
  NN \to NN  \nu \bar{\nu}}$  &
  $\quad 10^{19 \rrr}-10^{20}$\\
   \hline
\end{tabular}
\label{tab:slow}
\end{center}
\end{table*}

{\it Slow} neutrino reactions operate everywhere
in the core, particularly in the outer core
(and, hence, in low-mass stars). For matter consisting only of neutrons,
protons and electrons, they are
listed in Table \ref{tab:slow} (where $N$ is a nucleon, $n$ or $p$).
They are the modified Urca
process and $N\!N$-bremsstrahlung.
The modified Urca processes differ from their direct Urca counterparts
by an additional spectator nucleon $N$ required to ensure conservation of
momentum
and energy.
There are three bremsstrahlung processes ($nn$, $np$,
and $pp$) in $npe$ matter. There are other modified Urca and
bremsstrahlung processes in the presence of hyperons or quarks
(e.g., Yakovlev et al.\ 2001a).

The neutrino reactions are drastically affected by
baryon superfluidity
as reviewed by
Yakovlev et al.\ (1999, 2001a). 
When the temperature
$T$ drops much below the critical temperature $T_{\rm c}$
of a given baryon species $j$, the energy gap in the baryon energy
spectrum makes these baryons inactive, greatly
(as a rule, exponentially) suppressing all reactions
involving such baryons. For instance, a strong
superfluidity of protons in $npe$ matter suppresses
all Urca processes, but does not affect
neutron-neutron
bremsstrahlung.

While superfluidity suppresses the traditional processes,
it initiates a specific neutrino process
associated with {\it Cooper pairing of baryons}
(Flowers et al.\ 1976). 
In quasiparticle language, it may be described
as annihilation of two quasibaryons with the production of a neutrino pair.
This process is forbidden in nonsuperfluid systems, for which the
quasiparticles (the elementary excitations) are particles and holes.  Two
particles or two holes cannot annihilate because the total number of particles
is conserved, and the annihilation of particle and a hole is forbidden for
kinematical reasons.  With decreasing temperature, the process begins
to operate at
$T=T_{\rm c}$, produces the maximum emissivity at $T \sim 0.8\,T_{\rm c}$,
and becomes exponentially suppressed at $T \ll T_{\rm c}$.
For realistic density profiles $T_{\rm c}(\rho)$
at $T$ much below the maximum value of $T_{\rm c}(\rho)$,
the total neutrino luminosity of the star due to this
process cannot be too high: it can exceed the
luminosity provided by the modified Urca process
in a nonsuperfluid star at the same $T$ by up to two orders of magnitude.
The
Cooper pairing neutrino process operates in the core and the inner
crust.

Neutrino emissivities of many processes are model dependent, and
they have not been calculated with high precision.
For instance, the emissivities of the modified Urca
and nucleon bremsstrahlung processes in the
nonsuperfluid $npe$ matter are usually taken from
Friman \& Maxwell (1979), who used
the one-pion-exchange
Born approximation with phenomenological corrections.
Recently,
neutron-neutron bremsstrahlung
has been reconsidered by Hanhart et al.\ (2001),
van Dalen et al.\ (2003), and Schwenk et al.\ (2004)
using improved input for the nucleon-nucleon interactions, and
their results are in reasonable agreement with
those of Friman \& Maxwell (1979)
(with the phenomenological corrections included).
Among other recent papers, which have not been reviewed
by Yakovlev et al.\ (2001a),
let us mention the paper by Carter \& Prakash (2002) who
considered the renormalization of the axial weak interaction
constant by in-medium effects in dense matter. Gusakov
(2002) has studied the suppression of the modified Urca process
by the combined action of the single-state proton pairing
and the non-standard triplet-state neutron pairing
with the nodes of the gap on the neutron Fermi surface.
Jaikumar \& Prakash (2001) have analyzed neutrino
emission due to Cooper pairing of quarks.

Fortunately, cooling of neutron stars is
relatively insensitive to uncertainties in the neutrino emissivity
by a factor of 2--3 because
the emissivity is very temperature
dependent: such uncertainties are easily absorbed
by small variations of the internal temperature.

\subsection{Heat capacity}
\label{heat capacity}

The major contribution to the heat capacity of a neutron star comes from
the core, and it is
the sum of the heat capacities of various degenerate
constituents of the dense matter
(the contribution
of the crust is outlined, e.g, by
Gnedin et al.\ 2001).
The heat capacity per unit volume of
normal (nonsuperfluid) particle species $j$ is
$c_j=m_j^\ast p_j k_{\rm B}^2T/(3 \hbar^3)$,
where $p_j$ is the Fermi momentum, and $m_j^\ast$ is
the effective mass at the Fermi surface.
The main contribution to the heat capacity
of a nonsuperfluid core of neutrons, protons and electrons
comes from neutrons, while the contribution of
protons and electrons
is $\sim$25\% and $\sim$5\%, respectively (Page 1993).
The total thermal energy of a nonsuperfluid neutron star
is estimated as $U_T \sim 10^{48}\,T_9^2$ erg.

When $T$ drops below a critical temperature $T_{\rm c}$
of particle species $j$,
the heat capacity $c_j$ first jumps up discontinuously (by a factor of
2.2--2.4) but at $T \ll T_{\rm c}$ it becomes strongly suppressed
(as reviewed, e.g., by Yakovlev et al.\ 1999).
For instance, the heat capacity of an $npe$ neutron star core with
strongly superfluid neutrons and protons is determined by the
electrons, which are not superfluid, and it is about 20 times lower
than for a neutron star with a  nonsuperfluid core.

\section{COOLING THEORY}
\label{cooling theory}

\subsection{Basic Formalism}
\label{main equations}

Neutron stars are born very hot in supernova explosions,
with internal temperature $T \sim 10^{11}$ K,
but gradually cool down.
For about one minute following its birth, the star
stays in a special {\it proto-neutron star} state:
hot, opaque to neutrinos, and larger that an ordinary neutron star
(see, e.g., Pons et al.\ 2001
and references therein).
Later the star becomes
transparent to neutrinos generated
in its interior and transforms into an
ordinary neutron star. We consider the cooling during the
subsequent neutrino-transparent stage.
The cooling is realized via
two channels -- by neutrino emission from the entire stellar body
and by transport of heat
from the internal layers to the surface resulting
in the thermal emission of photons.

The internal structure of a neutron star
can be regarded as temperature-independent
(e.g., Shapiro \& Teukolsky 1983). 
For a given equation of state for dense matter, one can build a
family of neutron star models with different central densities $\rho_{\rm c}$
(hence, different masses and radii). Then one can simulate
the cooling of any model.

The general relativistic equations of thermal evolution
include the energy and flux equations obtained by
Thorne (1977). 
For a spherically symmetric star,
they are
\begin{equation}
    { {\rm e}^{-\lambda -2 \Phi} \over 4 \pi r^2 
    } \,
    { \partial  \over \partial  r}
    \left( {\rm e}^{2 \Phi} L_r \right)
    = -Q + Q_{\rm h}- {c_T \over {\rm e}^\Phi} \,
    {\partial T \over \partial t},  \quad {\rm and}
\label{therm-therm-balance}
\end{equation}
\begin{equation}
    {L_r \over 4 \pi \kappa r^2} =
    {\rm e}^{-\lambda - \Phi}
     {\partial \over \partial r} \left( T {\rm e}^\Phi \right),
\label{therm-Fourier}
\end{equation}
where $r$ is the radial coordinate (circumferential radius),
$Q$ is the neutrino emissivity,
$c_T$ is the heat capacity per unit volume,
$\kappa$ is the thermal conductivity,
and $L_r$ is the ``local luminosity'' defined as the
non-neutrino heat flux transported through a sphere of radius
$r$.
For completeness, we have introduced $Q_{\rm h}$, which represents the
rate of energy production
by  reheating sources (if any),
associated, for instance, with dissipation of rotational energy.
Furthermore, $\Phi(r)$ and $\lambda(r)$ are the metric functions
determined from a hydrostatic neutron-star model. The function $\Phi(r)$
specifies the gravitational redshift, while $\lambda(r)$ describes
the gravitational distortion of radial scales, ${\rm e}^{-\lambda}=
\sqrt{1-2Gm(r)/c^2r}$, where $m(r)$ is the gravitational mass
enclosed within a sphere of radius $r$. At the stellar
surface, $\Phi(R)=-\lambda(R)$.

In the outermost stellar
layers (roughly, as long as electrons are nondegenerate),
thermal conduction is radiative. Deeper in the crust, thermal
conductivity is provided by electrons (e.g., Potekhin et al.\ 1997),
while in the core
it is produced by electrons, neutrons, and
other baryons.

It is conventional (e.g., Gudmundsson et al.\ 1983) 
to separate
the stellar interior ($r< R_{\rm b}$)
and the outer heat-blanketing envelope
($R_{\rm b} \le r \le R$), where
the boundary radius $R_b$
corresponds to
a density $\rho_{\rm b} \sim 10^{10}$ g cm$^{-3}$
($\sim$ 100 meters under the surface).
The thermal structure of the blanketing envelope
is studied separately in the stationary,
plane-parallel approximation to relate
the surface temperature $T_{\rm s}$ (more generally,
the surface thermal luminosity, $L_\gamma$) to the temperature
$T_{\rm b}$ at the inner boundary of the envelope.
The $T_{\rm b}-T_{\rm s}$ relation is
used then as the boundary condition for solving
Eqs.\ (\ref{therm-therm-balance}) and (\ref{therm-Fourier})
at $r<R_{\rm b}$.

The main goal of the cooling theory is to calculate
{\it cooling curves}, $T_{\rm s}^\infty(t)$ (or $L_\gamma^\infty(t)$),
to be compared with observations.
One can distinguish three main cooling stages:

({\it i}) The initial {\it thermal relaxation
stage} lasts for $t \la 10$--100 yrs;
the crust remains thermally decoupled from the core, and the surface
temperature reflects the thermal state of the crust
(Lattimer et al.\ 1994,
Gnedin et al.\ 2001). 

({\it ii}) The {\it neutrino cooling stage} (neutrino luminosity
$L_\nu \gg L_\gamma$)
lasts for $t \la 10^5$
yr;
the cooling is
produced by
neutrino emission from the stellar interior
(mainly from the core), while the surface temperature
adjusts to the internal one.

({\it iii}) During the final {\it photon cooling stage} ($L_\nu \ll L_\gamma$
$t \ga 10^5$ yr) the star cools via photon emission
from the surface, and the evolution of the internal temperature
is governed by the radiation from the stellar surface, and hence
it is sensitive to properties of the outer parts of the star.

After the thermal relaxation, the redshifted
temperature
$ T_i(t)= T(r,t)\; {\rm e}^{\Phi(r)}$
becomes constant throughout the stellar interior.
Then Eqs.\ (\ref{therm-therm-balance}) and (\ref{therm-Fourier})
reduce to the equation of global thermal balance
(Glen \& Sutherland 1980),
\begin{equation}
   C(T_i) \,
     {{\rm d} T_i \over {\rm d} t}  =
      - L_\nu^\infty ( T_i ) + L_{\rm h}^\infty - L_\gamma^\infty (T_{\rm s}),
\label{therm-isotherm}\\
\end{equation}
\begin{equation}
   L^\infty_\nu ( T_i) =
     \int
     {\rm d}V \, Q (T) \, {\rm e}^{2 \Phi},
\,\,{\rm and}\,\,
     L^\infty_{\rm h}  =
     \int
     {\rm d}V \, Q_{\rm h} \, {\rm e}^{2 \Phi},
\quad
     C(T_i) =
      \int {\rm d}V  c_T(T),
\label{therm-C}
\end{equation}
where ${\rm d}V=4 \pi r^2 {\rm e}^\lambda \;{\rm d}r$ is the element of
proper volume,
$C$ is the total stellar heat capacity,
$L^\infty_\nu$ is the total neutrino
luminosity (for a distant observer), and
$L_{\rm h}^\infty$ is the total reheating power.

\subsection{Heat-Blanketing Envelope:
$T_{\rm s}-T_{\rm b}$ Relation}
\label{TsTbrelation}

An accurate $T_{\rm s}$--$T_{\rm b}$ relation
for a nonmagnetic heat-blanketing iron envelope
at $T_{\rm s} \ga 2 \times 10^5$ K was obtained
by Gudmundsson et al.\ (1983): $T_{\rm b}=1.288 \times 10^8\,
(T_{\rm s6}^4/g_{14})^{0.455}$ K, where $g_{\rm 14}$
is the surface gravity $g=GM{\rm e}^{-\Phi(R)}/R^2$
in units of $10^{14}$ cm s$^{-2}$, and $T_{\rm s6}=T_{\rm s}/
10^6$ K. The
thermal insulation is actually
provided by a relatively thin layer of the  outer part of the blanketing
envelope
(where the electrons are mildly degenerate);
this layer becomes thinner as
$T_{\rm s}$ decreases.

Potekhin et al.\ (1997) extended these results to
lower temperatures and took into account the possible
presence of a thin surface layer composed of light
elements, primarily H and He, which could be the result of accretion.
The thermal conductivity of the light-element plasma
is higher than that of an iron plasma,
thereby
increasing $T_{\rm s}$ for a given $T_{\rm b}$.
The effect is determined by the total mass $\Delta M$ of the light-element
envelope, and is insensitive to whether the accreted matter is
H or
He. Because of beta captures and pycnonuclear burning
of light elements at densities $\rho
\ga
10^{10}$ g cm$^{-3}$ the mass of
a light-element envelope is limited to
$\Delta M \la 10^{-7}$ \msun.

A magnetic field influences the
thermal conductivity of the surface layers and, in particular, makes
it anisotropic. This affects the $T_{\rm s}-T_{\rm b}$ relation
(e.g., Potekhin et al.\ 2003 and references
therein). The effects are twofold.

({\it 1}) {\it Classical} effects
are produced by electron cyclotron motion perpendicular to the direction
of the
magnetic field. These effects can strongly reduce the electron
thermal conductivity across the magnetic field but
they do not affect the conductivity along the field.
They are especially important
near the magnetic equator, where the
field is tangential to the surface and the heat is carried
away from the
stellar
interior
across the field lines.
Such equatorial regions conduct heat less well, which
lowers the local effective
temperature for a given $T_{\rm b}$.

({\it 2}) {\it Quantum} effects are associated with the quantization
of electron motion into Landau levels. They
may strongly modify the conductivities along and
across the field lines. If the magnetic field is so strong
that the majority of electrons occupy the lowest Landau
level, the quantum effects enhance the longitudinal
thermal conductivity of degenerate electrons.
This effect is most pronounced
near the magnetic poles, where the field is normal to the surface and
heat propagates
along the field lines. The quantum effects
increase the local surface temperature for a given
$T_{\rm b}$.

The total photon luminosity of the star is obtained by integrating
the local radiated flux over the entire stellar surface
(Potekhin \& Yakovlev 2001, Potekhin et al.\ 2003).
For instance, at $T_{\rm s} \ga 3 \times 10^5$ K
the dipole magnetic field affects the luminosity if
$B \ga 3 \times 10^{10}$ G. As long as
$B \la 3 \times 10^{13}$ G, the equatorial
decrease of the heat transport dominates, and the
luminosity is lower than at $B=0$. For higher $B$,
the polar increase of the heat transport dominates,
and the magnetic field increases
the photon luminosity.
At $B \ga 10^{12}$ G the magnetic poles
become much hotter than the equator. However,
due to the gravitational lensing
effect,
 the
flux of thermal radiation is almost independent of
the direction of observation (Page 1995,
Potekhin \& Yakovlev 2001).

The effects of a magnetic field
are most important
in the outer part of the heat-blanketing envelope
and they weaken with increasing $\rho$.
This has two consequences. {\it First}, in a hot neutron star
($T_{\rm s} \ga 3 \times 10^6$ K), the most important heat-insulating
layer lies deep in the blanketing envelope.  As a result,
the $T_{\rm s}-T_{\rm b}$ relation is weakly affected
by the magnetic field, and for very high $T_{\rm s}$
it  {\it converges to the field-free result}.
Thus even a very strong field cannot change the thermal
state of a hot neutron star.
{\it Second}, the temperature
distribution in the interior ($\rho>\rho_{\rm b}$)
may be regarded as spherically symmetric. This justifies
the use of the
purely radial equations (\ref{therm-therm-balance}) and
(\ref{therm-Fourier}) in the internal region.
In the presence of strong magnetic fields,
it is reasonable to shift $\rho_{\rm b}$ to the neutron drip density,
$4 \times 10^{11}$ g cm$^{-3}$
(Potekhin \& Yakovlev 2001). 

Recently the effects of light element envelopes and
magnetic fields have been reconsidered by Potekhin et al.\ (2003)
and Yakovlev et al.\ (2004b). Potekhin et al.\ (2003)
have also studied the combined effects of light elements
and magnetic fields. However, the available
$T_{\rm s}-T_{\rm b}$
relations of cold ($T_{\rm s} \la 10^5$ K) and/or
strongly magnetized ($B \ga 10^{13}$ G) neutron stars
are still far from perfect because of the problems
of calculating ionization equilibrium, the equation of state, and the thermal
conductivity of cold and/or strongly magnetized plasmas.

The heat propagation time $t_{\rm th}$ through the blanketing envelope
depends on many factors. For $T_{\rm s} \sim 10^6$ K and
$\rho_{\rm b} \sim 10^{10}$ g cm$^{-3}$ one has $t_{\rm th} \sim 1$ yr
(e.g., Ushomirsky \& Rutledge 2001). In a cooler star
$t_{\rm th}$ is shorter due to the
higher thermal conductivity. 

\subsection{Physical Properties that Determine Cooling }
\label{cooling regulators}

Cooling of neutron stars is affected by many factors, of which the
most important are:

({\it a}) The rate of neutrino emission from
the interior of a
neutron star.

({\it b}) The heat capacity
in the stellar interior.

({\it c}) The first two items depend in turn on the composition of matter in
the stellar interior, and the gross stellar structure depends on the equation
of state of matter at high densities.

({\it d}) Superfluidity of matter can have a dramatic effect on
neutrino emission and the heat capacity.\footnote{
In this way, as pointed out by Page \& Applegate
(1992), cooling neutron stars may serve as
thermometers for measuring
critical temperatures for nucleon superfluidity in the interiors of neutron
stars.}

({\it e}) The thermal conductivity, especially in the heat blanketing
envelope, is a crucial ingredient, because
it
determines the relationship
between $T_{\rm s}$ and $T_{\rm b}$.  In particular, the thermal conductivity
is sensitive to the presence of a light-element
layer and
magnetic fields
in the blanketing envelope.

({\it f})
Possible reheating mechanisms such as frictional dissipation
of rotational energy
may be especially important in cold and old neutron stars.
For simplicity, we shall neglect them
in cooling calculations (setting $Q_{\rm h} \equiv 0$), but
we shall summarize them in Sect.\ \ref{old neutron stars}.

\section{COOLING OF NEUTRON STARS WITH NUCLEON CORES}
\label{nucleon cores}

\subsection{Physics Input}
\label{input}

We start
with the simplest composition
for
a neutron star core, just neutrons,
protons and electrons. Illustrative cooling curves are
calculated with our fully relativistic
nonisothermal cooling code (Gnedin et al.\ 2001).
In the stellar core, we use
a stiff phenomenological
equation of state proposed by Prakash et al.\ (1988)
(their model I for the symmetry energy and the
model of the bulk energy which gives for
the compression modulus of saturated
nuclear matter $K=240$ MeV).
The parameters for neutron star models with several masses
are given in Table \ref{tab:NSmodels}.
Along with the values of $M$ and $R$ we present
the central density $\rho_{\rm c}$, the total mass of the inner and outer
crusts
$M_{\rm crust}$,
the total width of
these crusts $\Delta R_{\rm
crust}$
(defined as $R-R_{\rm core}$, where
$R_{\rm core}$ is the radius
of the crust-core interface),
the 
mass $\Delta M_{\rm D}$ and radius $R_{\rm D}$
of the central
core,
where the direct Urca process can occur.
The most massive stable neutron star for this equation of state
has $M_{\rm max}=1.977\,\msun$
and $\rho_{\rm c}=2.578 \times 10^{15}$
g cm$^{-3}$.
For the given equation of state,
the direct Urca process is
allowed at densities
$\rho \geq \rho_{\rm D}=7.851 \times 10^{14}$ g cm$^{-3}$.
The mass of the
star with $\rho_{\rm c}=\rho_{\rm D}$
is $M=M_{\rm D}=1.358\,\msun$.

\begin{table*}[!t]
\caption{Neutron star models}
\begin{center}
\begin{tabular}{||lrrllll||}
\hline\hline
       $M$  & $R~~~$  & $\rho_{\rm c}$ ($10^{14}$
    & $M_{\rm crust}$ & $\Delta R_{\rm crust}$
    & $\Delta M_{\rm D}$ & $R_{\rm D}$  \\
   ($M_\odot$) & (km) & g cm$^{-3}$)   & ($M_\odot$)
    & (km)  & ($M_\odot$) & (km) \\
\hline
 1.1      & 13.20 &  6.23 & 0.069 & 1.98 & \ldots & \ldots \\
 1.2      & 13.13 &  6.80 & 0.063 & 1.77 & \ldots & \ldots \\
 1.3      & 13.04 &  7.44 & 0.057 & 1.58 & \ldots & \ldots \\
 1.358$^a$& 12.98 &  7.85 & 0.054 & 1.48 & 0.000  & 0.00 \\
 1.4      & 12.93 &  8.17 & 0.052 & 1.40 & 0.023  & 2.40 \\
 1.5      & 12.81 &  9.00 & 0.049 & 1.26 & 0.137  & 4.27 \\
 1.6      & 12.64 & 10.05 & 0.042 & 1.10 & 0.306  & 5.51 \\
 1.7      & 12.43 & 11.39 & 0.035 & 0.96 & 0.510  & 6.41 \\
 1.8      & 12.16 & 13.22 & 0.030 & 0.84 & 0.742  & 7.10 \\
 1.9      & 11.73 & 16.33 & 0.023 & 0.69 & 1.024  & 7.65 \\
 1.977$^b$& 10.75 & 25.78 & 0.011 & 0.45 & 1.400  & 7.90 \\
\hline
\end{tabular}
\begin{tabular}{l}
 \small{$^a$  Threshold configuration for the direct Urca process} \\
 \small{$^b$  Maximum-mass stable neutron star }
\end{tabular}
\label{tab:NSmodels}
\end{center}
\end{table*}

The cooling code includes the effects of nucleon
superfluidity
of three types: singlet-state ($^1$S$_0$) pairing
of free neutrons in the inner crust
(with a critical temperature
$T_{\rm c}=T_{\rm cns}(\rho)$); $^1$S$_0$ proton pairing
in the core ($T_{\rm c}=T_{\rm cp}(\rho)$);
and triplet-state ($^3$P$_2$) neutron pairing
in the core ($T_{\rm c}=T_{\rm cnt}(\rho)$).
The $T_{\rm c}(\rho)$ dependence
has been parameterized by simple equations (e.g., Yakovlev
et al.\ 2002b);
the models for $T_{\rm c}(\rho)$ are qualitatively the same
as those obtained in a number
of microscopic calculations.

\begin{figure}[!t]        
\begin{center}
\leavevmode \epsfxsize=\textwidth
 \epsfbox[50 180 550 460]{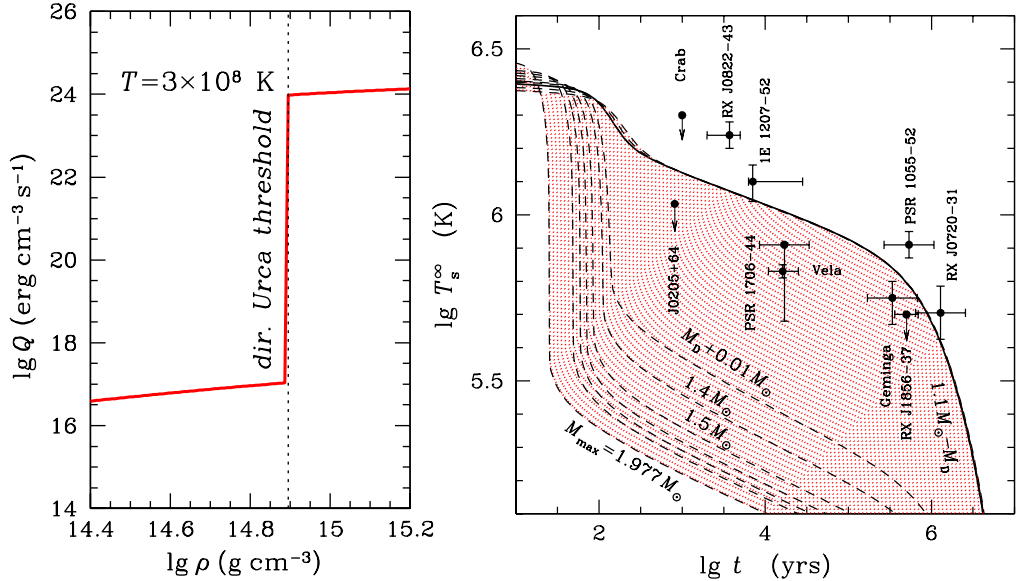}
\end{center}
\caption{{\it Left}: Density profile of neutrino emissivity
throughout the core of a nonsuperfluid neutron star at
$T=3 \times 10^8$ K. {\it Right}: Cooling curves for
nonsuperfluid neutron stars of several masses compared with
observations; points lying within the shaded region may formally be explained
by models of cooling nonsuperfluid neutron stars with the given equation of
state.}
\label{fig:nosf}
\end{figure}

\begin{figure}[!th]        
\begin{center}
\leavevmode \epsfysize=6.5cm
 \epsfbox[50 180 465 410]{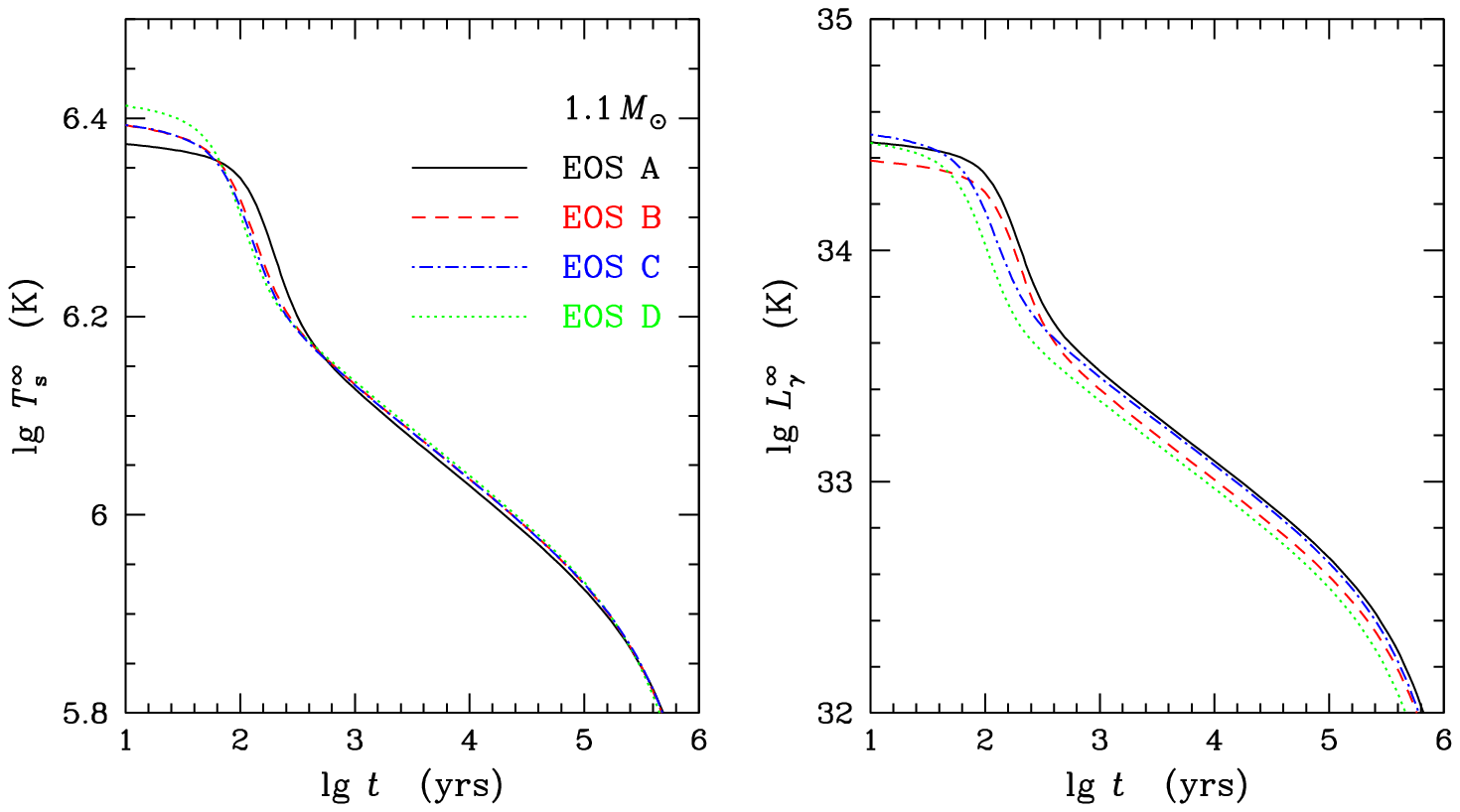}
\end{center}
\caption{
Surface temperatures
({\it left}) and photon luminosities ({\it right})
of nonsuperfluid $1.1 \, \msun$ neutron stars
for four high-density equations of state (see text).}
\label{fig:universal}
\end{figure}

\subsection{Nonsuperfluid Stellar Models}
\label{nosf}

In this subsection we neglect the effects
of light-element surface envelopes and surface magnetic fields.
For nonsuperfluid neutron stars,
we have {\it two} well-known
cooling regimes, {\it slow} and {\it fast} cooling
due to slow and fast neutrino emission
(Sect.\ \ref{neutrinos}) as illustrated in Fig.\ \ref{fig:nosf}.
The left panel shows the profile
of neutrino emissivity in the core at $T=3 \times 10^8$ K
with a jump by 7 orders of magnitude at the direct Urca threshold.
The right panel shows cooling curves of neutron stars with
several masses $M$: 1.1, 1.2,
1.3 \msun , $M_{\rm D}$,
$M_{\rm D}+0.01\, \msun$, 1.4, 1.5, 1.6, 1.7, 1.8 \msun ,
and $M_{\rm max}$.

{\it Slow cooling} occurs in low-mass
stars ($M< M_{\rm D}$) via neutrino emission produced mainly by
the modified Urca process. The cooling
curves are almost the same
for all $M$ from about \msun\
to $M_{\rm D}$ (Page \& Applegate 1992).
The surface temperature stays high, $T_{\rm s}^\infty \ga 10^6$ K,
for $t \la 10^5$ yrs.
A simple estimate from
the thermal balance equation, Eq.\ (\ref{therm-isotherm}),
gives an approximate slow-cooling law during the neutrino
cooling stage: $t_{\rm slow} \sim 1$ yr$/T_{i9}^6$
(see, e.g., Pethick 1992). The internal temperature
drops to $T_i \sim 1.5 \times 10^8$ K in $t=10^5$ yrs.
These nonsuperfluid models of
cooling by the  modified Urca process
cannot explain the observational limits of some sources,
first of all, PSR J0205+6449,
Vela, and PSR B1706--44 (too cold),
as well as RX J0822--4300 and PSR B1055--52 (too hot).
The data seem to require both slower and faster cooling.

The {\it fast cooling} occurs at $M -M_{\rm D} \ga 0.01 \, \msun$
via the powerful direct Urca process.
The cooling curves are again not too sensitive to the stellar mass.
These stars are much colder
($T_{\rm s}^\infty \la 3 \times 10^5$ K
for $t \sim 10^4$ yrs) than the slow-cooling ones.
An estimate from Eq.\ (\ref{therm-isotherm})
during the neutrino cooling stage now yields
$t_{\rm fast} \sim 1$ min$/T_{i9}^4$, giving
$T_i \sim 10^7$ K for $t=200$ yrs.

The {\it transition} from the slow to fast cooling takes
place in a very narrow range of $M$
because of the huge difference in the emissivities
of the modified and direct Urca processes, and the sharp threshold
of the direct Urca process
(left panel of Fig.\ \ref{fig:nosf}).
On the cooling diagram
some sources (in particular, Vela,
PSR B1706--44, 
Geminga, RX J1856.4--3754)
fall exactly in this transition zone, and therefore
could be explained if they had almost the same mass.
This unlikely assumption can be avoided by
including the effects of nucleon superfluidity
(see Sect.\ \ref{psf}).

Let us stress the {\it universality} of the cooling
curves (Page \& Applegate 1992).
The curves for low-mass stars are insensitive not only
to the values of $M$ (as long as $M<M_{\rm D}$),
but also to the equation of state at high densities. The effect is illustrated
in Fig.\ \ref{fig:universal}, where we show
the cooling of 1.1 \msun\ stars,
with four equations of state. Model A
is our basic equation of state (Sect.\ \ref{input}), while
B, C, and D are modifications of the
equations of state of Prakash et al.\ (1988) with
the simplified form of the symmetry energy
suggested by Page \& Applegate (1992).
Specifically, models B, C, and D correspond to three values of the
compression modulus of saturated nuclear matter:
$K$=180, 240, and 120 MeV, respectively. They
are examples of moderate, stiff, and soft equations of state (yielding
$M_{\rm max}$=1.730, 1.942,
and 1.461 \msun). Although the equations of state and the stellar models are
different, the cooling curves of low-mass stars are almost the same,
especially if one plots $T_{\rm s}^\infty(t)$
(left panel) rather than $L_\gamma^\infty(t)$
(right panel). This universality is easily explained
from Eq.\ (\ref{therm-isotherm}):
the cooling rate at the neutrino
cooling stage is proportional to $L_\nu^\infty/C$,
the ratio of the total neutrino luminosity and heat capacity.
While
both $L_\nu^\infty$ and $C$
depend on the stellar model,
their ratio is almost
model-independent. Thus, we have actually
one universal {\it standard basic cooling curve} for all models
with $\msun \la M<M_{\rm D}$.
This curve is plotted in Fig.\ \ref{fig:observ}.
The cooling curves of high-mass
stars are also similar (for the same reasons).
For instance, the
curves of maximum-mass
stars
with equations of state A--D are almost identical.

\subsection{Effects of Magnetic Fields and Light-Element Envelopes}
\label{magnetic fields}

\begin{figure}[!t]        
\begin{center}
\leavevmode \epsfxsize=\textwidth
 \epsfbox[50 190 550 380]{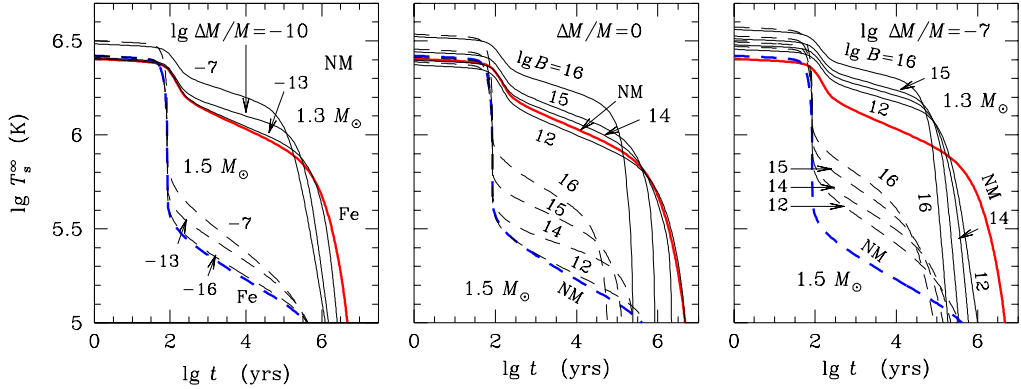}
\end{center}
\caption{Cooling of nonsuperfluid neutron stars with $M=1.3\,M_\odot$
(solid lines)
and $1.5\,M_\odot$ (dashed lines).
{\it Left}: nonmagnetic (NM) stars with different
amounts $\Delta M$ of light elements
in the blanketing envelopes; thick curves refer to
nonaccreted (Fe) envelopes.
{\it Middle}:
stars with nonaccreted envelopes
and
dipole surface magnetic fields (the curves are
labeled by $\lg B_{\rm p}$,
and thick lines refer to
$B=0$).
{\it Right}: same as in the
middle
panel but with a fully
accreted envelope.
\label{tript}}
\end{figure}

Now we discuss the effects
of surface magnetic fields and light-element (accreted)
envelopes (Sect.\ \ref{TsTbrelation}).
We shall follow the considerations of Potekhin et al.\ (2003)
(also see their paper for references to earlier work).
Figure \ref{tript} shows slow and fast cooling
of nonsuperfluid 1.3 \msun\ (solid lines) and 1.5 \msun\
(dashed lines) stars.

The left panel illustrates the effects
of accreted envelopes in nonmagnetized stars.
We present the cooling curves for some
values of $\Delta M$, the mass of relatively light
elements (H, He, C, or O) in the heat-blanketing envelopes.
The fraction of accreted mass $\Delta M/M$
varies from 0 (nonaccreted envelopes) to $\sim 10^{-7}$
(fully accreted envelopes;
see Sect.\ \ref{TsTbrelation}).

During the neutrino cooling stage, the internal
stellar temperature $T_{\rm b}$ is determined by neutrino emission,
and is insensitive to the physics of the heat-blanketing
envelope. The surface temperature
adjusts to $T_{\rm b}$ according to
the $T_{\rm s}$--$T_{\rm b}$ relation.
Since accreted envelopes conduct heat better than ones composed of heavier
elements, the surface
temperature of a star with an accreted envelope is noticeably higher.
Even an amount of accreted matter as small as
$\Delta M/M \sim 10^{-13}$ can appreciably change
the cooling.
The cooler the star, the smaller $\Delta M$
which yields the same cooling
as the fully accreted blanketing envelope.
The reason for this is that the greatest contribution to the difference
between surface and interior temperatures occurs in a thin layer where
matter is partially degenerate.
The cooler the star, the closer
this layer is to the surface, and the smaller the amount of accreted matter
is
needed to change the composition of this layer.

For $t \ga 10^5$ yrs the star enters
the photon cooling stage
when the cooling is governed by $T_{\rm s}$ (Sect.\ \ref{main equations}).
An accreted envelope leads then to
faster cooling (more rapid fall
of $T_{\rm s}$ with time).
Thus, the effects of light elements
on the surface temperature during the
neutrino and photon cooling stages are opposite. This reversal
of the effect of light elements when passing from one stage to the other
has a straightforward physical explanation, and is well known.

The middle panel of Fig.\ \ref{tript} displays the effect
of a dipolar magnetic field on the cooling
of neutron stars with nonaccreted envelopes. Let us remark
that in this case  $T_{\rm s}$ refers to
the average surface temperature (Sect.\ \ref{observ}).
We present the
cooling curves for several strengths
of the magnetic field at the poles up
to $B_{\rm p}=10^{16}$ G. The cooling curves of nonmagnetic neutron stars
are shown as thick lines.
For simplicity, the magnetic field is assumed to be independent of
time.

A magnetic field $B_{\rm p} \la 10^{13}$~G makes
the blanketing envelope of a warm ($1.3\,M_\odot$) neutron star
less
conducting to heat
(Sect.\ \ref{TsTbrelation}).
This lowers $T_{\rm s}$ during the neutrino cooling
stage  and increases $T_{\rm s}$ during the photon cooling stage,
producing another reversal effect.
By contrast, a stronger magnetic field with $B_{\rm p} \gg
10^{13}$
G makes the blanketing envelope a better conductor of heat, which
increases $T_{\rm s}$ during the neutrino cooling stage
and lowers $T_{\rm s}$ during the photon cooling stage (another
reversal).
A field $B_{\rm p} \sim 10^{13}$ G has almost no effect
on the cooling.
The fast cooling of cooler magnetized (1.5 $M_\odot$) neutron stars
is somewhat different:
a magnetic field $B_{\rm p} \la 10^{13}$ G
has almost no effect on $T_{\rm s}$,
while
fields $B_{\rm p} \ga 10^{13}$ G affect the
cooling much more than
for slow-cooling neutron stars.

The right panel of Fig.\ \ref{tript} shows cooling curves for neutron stars
with fully accreted envelopes and the same dipole magnetic fields as in the
middle panel.  For a star with $B_{\rm p} \la 10^{15}$ G during the neutrino
cooling stage, the effect of the accreted envelope is stronger than the effect
of the magnetic field.  For higher $B_{\rm p}$, the magnetic effect dominates;
the accreted envelope produces a rather weak additional rise in $T_{\rm s}$.

The main outcome of these studies is that even ultrahigh magnetic fields do
not change the average surface temperatures of young and warm neutron stars as
much as an accreted envelope can.  At the same time, strong fields
induce a strongly nonuniform surface temperature distribution
(see, e.g.,
Potekhin et al.\ 2003).

\subsection{Proton Superfluidity and
Three Types of Cooling Neutron Stars}
\label{psf}

The considerations above show that the effects of magnetic fields
and accreted envelopes in nonsuperfluid neutron stars
cannot reconcile theory and observation.
Thus, we turn to the cooling of {\it superfluid} neutron stars.
For simplicity, in Sects.\ \ref{psf} and \ref{pnsf}
we consider nonmagnetic neutron stars without accreted envelopes.

\begin{figure}[!t]        
\begin{center}
\leavevmode \epsfxsize=12cm
 \epsfbox[50 180 570 470]{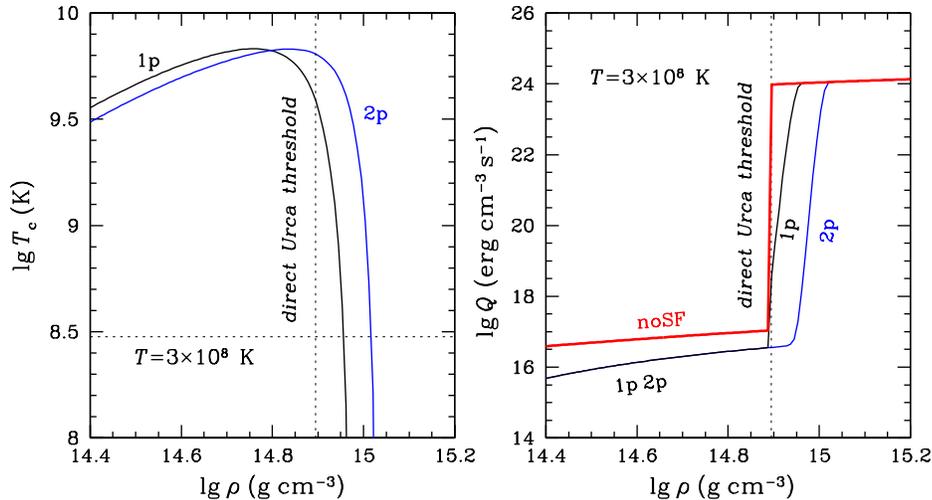}
\end{center}
\caption{{\it Left:} Superfluid transition temperature versus density for two
models (1p and 2p)
for proton superfluidity
in the
neutron star
core. {\it Right:} Neutrino
emissivity profiles in the core at $T=3\times10^8$ K
for nonsuperfluid matter (noSF)
and for matter with superfluid protons
(models
1p or 2p).}
\label{fig:psf}
\end{figure}

The observations can be explained by the cooling of superfluid neutron stars
assuming that
{\it proton superfluidity
is rather strong} at $\rho \la \rho_{\rm D}$, while
the $^3$P$_2$ {\it neutron superfluidity is rather weak}.
We start with the effects of proton superfluidity
(Kaminker et al.\ 2001)
and neglect neutron pairing.
We take two typical models of proton superfluidity,
1p and 2p. The model critical temperatures
$T_{\rm cp}(\rho)$ are displayed in the left panel of Fig.\ \ref{fig:psf}.
The resulting neutrino emissivity in the stellar core at
$T=3 \times 10^8$ K is
shown in the right panel.

The effects of
proton superfluidity are seen to be twofold.
First, superfluidity reduces the neutrino
emission in the outer core by strongly suppressing the
modified Urca and even the direct Urca process at not too high $\rho$.
Consequently,
neutron-neutron bremsstrahlung (Table \ref{tab:slow}), which is weaker,
becomes the leading neutrino emission mechanism.
Second, proton superfluidity gradually dies out
with increasing $\rho$, smoothly
removing the reduction
of fast neutrino emission.
This broadens the
direct Urca threshold, creating
{\it a finite transition zone}
 at densities $\rho_{\rm s} \la \rho \la
\rho_{\rm f}$
between the regions with slow neutrino emission ($\rho \la
\rho_{\rm s}$)
and rapid neutrino emission
($\rho \ga \rho_{\rm f})$.
For model 2p, superfluidity extends deeper into the stellar core and
shifts the transition zone to higher densities.
The direct Urca threshold can also
be broadened by the
thermal effects
and by
magnetic fields (Baiko \& Yakovlev 1999) but these
effects are usually weaker than the broadening provided
by superfluidity (Yakovlev et al.\ 2001a).

\begin{figure}[!th]        
\begin{center}
\leavevmode \epsfxsize=12cm
 \epsfbox[40 180 535 500]{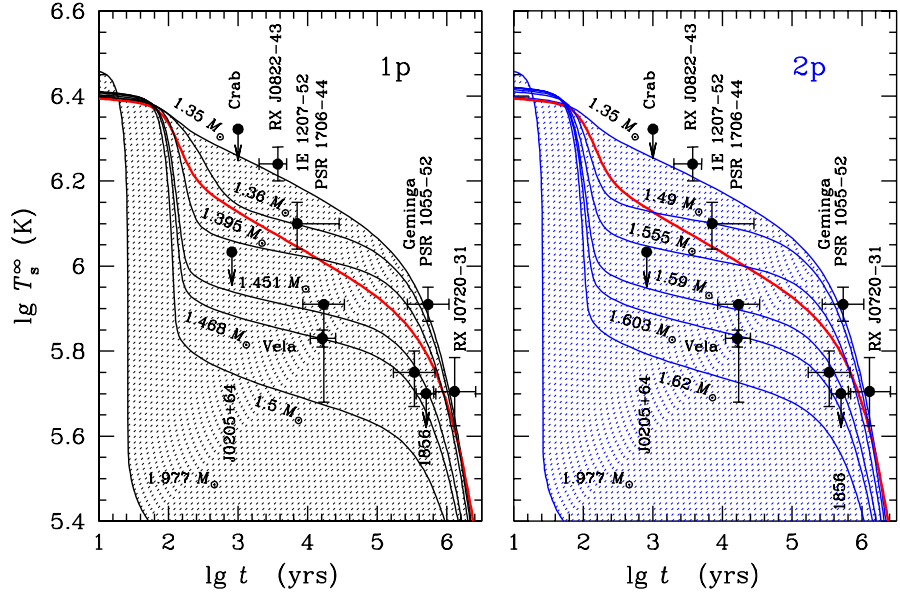}
\end{center}
\caption{Cooling of neutron
stars
of different masses with either
model 1p ({\it left}) or model 2p ({\it right}) for proton superfluidity
in the cores together with observations of isolated neutron stars.
Any point lying within the shaded regions can be explained by the given
models for some value of the neutron star mass.}
\label{fig:1p2p}
\end{figure}

The cooling curves of neutron stars of different masses
with proton superfluidity models 1p and 2p are plotted
in the left and right panels of Fig.\ \ref{fig:1p2p}, respectively.
We see that proton superfluidity
leads to {\it three}
characteristic
types of cooling neutron stars
(Kaminker et al.\ 2002).

{\it Low-mass} neutron stars, with $\rho_{\rm c} < \rho_{\rm s}$,
have weaker neutrino emission
than low-mass nonsuperfluid neutron stars, and they form
a class of {\it very slowly cooling neutron stars}.
Their cooling curves are almost universal
(as in Sect.\ \ref{nosf}): they are
independent of
the
stellar
mass, the model for proton superfluidity,
and the equation of state
in the stellar core.
These cooling curves lie above
the basic standard cooling curve, and can explain the observations
of RX J0822--4300 and PSR B1055--52.
Thus we can treat these two sources as low-mass neutron stars.
With some reservations, RX J0720.4--3125
may also be attributed to this class (although, given the large
observational
uncertainties,
it may belong to the class of cooler, medium-mass stars).

{\it High-mass, rapidly cooling} neutron stars, with
$\rho_{\rm c} \ga \rho_{\rm f}$,
cool mainly via fast neutrino emission
from the inner core. The cooling curves are again
almost independent of $M$, equation of state and model for proton
superfluidity, and they are actually the same as for high-mass nonsuperfluid
stars. All observed isolated neutron stars are much warmer than these
models.

{\it Medium-mass} neutron stars ($\rho_{\rm s} \la \rho_{\rm c} \la \rho_{\rm
f}$) show cooling which is {\it intermediate}
between very slow and fast; it
depends on $M$, the equation of state and proton
superfluidity. Roughly, the masses of these stars range
from $M_{\rm D}$ to 1.55 \msun\ for the 1p superfluidity model
and from 1.4 to 1.65 \msun\ for the 2p superfluidity model.
By varying $\rho_{\rm c}$ from
$\rho_{\rm s}$ to $\rho_{\rm f}$ we obtain a family
of cooling curves which fill the (shaded) space between
the curves of low-mass and high-mass stars. Then we
can select those curves which explain the observations
and thus attribute
masses to the sources.
This ``weighing of neutron stars'' suggested
by Kaminker et al.\ (2001) depends on a proton superfluidity model
(Fig.\ \ref{fig:1p2p})
as well as on the
equation of state
and the composition of matter in the core (see, e.g.,
Kaminker et al.\
2002), which
determine the position of the direct Urca threshold.
We can treat 1E 1207--52, Vela, PSR B1706--44, 
Geminga,
and RX J1856.4--3754 as medium-mass neutron stars.

\begin{figure}[!t]        
\begin{center}
\leavevmode \epsfxsize=12cm
 \epsfbox[40 180 575 470]{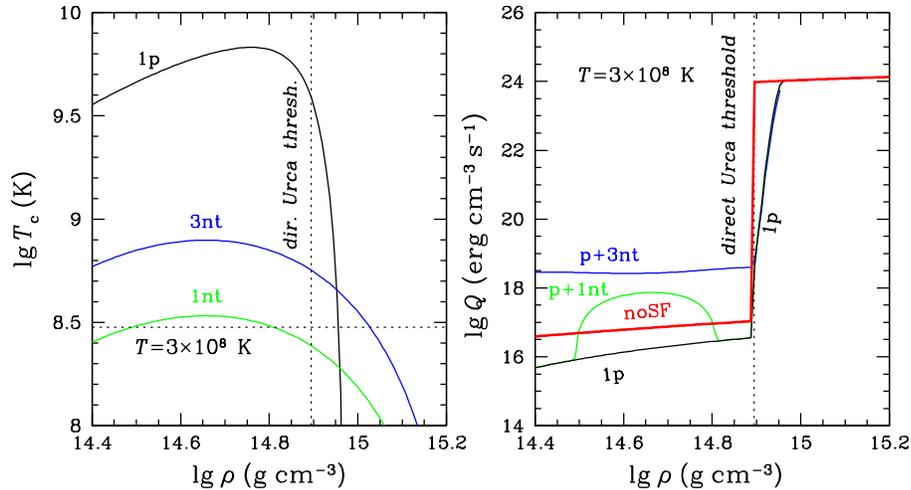}
\end{center}
\caption{{\it Left:} Superfluid transition temperatures as a function of
density for protons
(model 1p) and neutrons (models 1nt and 3nt for triplet-state pairing)
in the core of a neutron star. {\it Right:}
Neutrino emissivity as a function of density at $T=3\times10^8$ K
in nonsuperfluid matter (noSF) and
in the presence of proton superfluidity (model 1p) and neutron
superfluidity (models
1nt and
3nt).}
\label{fig:pnsf}
\end{figure}

\begin{figure}[!t]        
\begin{center}
\leavevmode \epsfxsize=\textwidth
 \epsfbox[20 150 480 420]{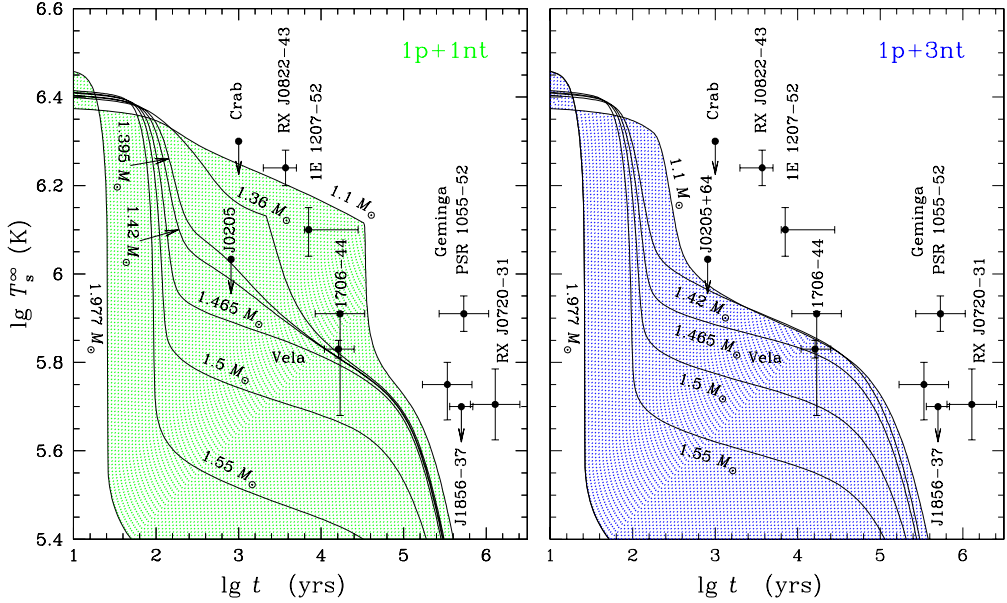}
\end{center}
\caption{Cooling of neutron stars of different masses with models
1p for proton superfluidity and either 1nt
 ({\it left}) or 3nt ({\it right}) for neutron superfluidity
in the cores together  with observations of isolated neutron stars.
Any point within the shaded regions may be explained by
the given cooling models for
some value of the stellar mass.}
\label{fig:1pnt}
\end{figure}

\subsection{Mild Neutron Pairing in the Core Contradicts Observation}
\label{pnsf}

We now investigate the effect of  $^3$P$_2$ neutron pairing in the stellar
core
(e.g., Kaminker et al.\ 2001, 2002; Yakovlev et al.\ 2002a).
Microscopic theories predict this pairing to be
weaker than the proton one.
Two models for $T_{\rm cnt}(\rho)$
(1nt and 3nt) are presented in the left panel
of Fig. \ref{fig:pnsf}, with
maximum $T_{\rm cnt}^{\rm max} \approx 3.4\times 10^8$ and
$8 \times 10^8$ K, respectively.
The appearance of such superfluidity induces the strong
neutrino emission due to Cooper pairing of neutrons
in the outer core as shown
in the right panel of Fig.\ \ref{fig:pnsf}.

Theoretical cooling
curves of neutron stars with model 1p for proton superfluidity
and with either model 1nt or model 3nt for neutron superfluidity are shown
in Fig.\ \ref{fig:1pnt}.
Before the onset of neutron superfluidity the curves
are the same as for models with proton superfluidity
alone (Fig.\ \ref{fig:1p2p}). After the onset,
cooling is strongly accelerated by neutrino emission
due to the Cooper pairing of neutrons.
For model 3nt, neutron superfluidity
is stronger than for model 1nt and appears earlier, producing
faster cooling (cf.\ right and left
panels of Fig.\ \ref{fig:1pnt}).
In any case, the fast cooling predicted by these models is in conflict with
observations of many sources.

In fact any {\it mild} neutron superfluidity in the stellar core with a
realistic $T_{\rm cnt}(\rho)$ profile and $T_{\rm cnt}^{\rm max}\sim (2 \times
10^8-2\times 10^9)$ K contradicts observations of at least some hotter and
older objects (independently of the proton pairing) and {\it should be
rejected} on these grounds (Yakovlev et al.\ 2004a). Neutron superfluidity
with a smaller $T_{\rm cnt}^{\rm max}$ would come into play only in the late
stages of neutron star evolution and has no effect on the cooling of
middle-aged stars. It is interesting that recent calculations by Schwenk \&
Friman (2004) indicate that the medium-induced 
one-pion-exchange interaction
(acting in second order)
greatly reduces
triplet-state pairing of neutrons in neutron star cores.
This is in line with the above conclusion that
triplet-state neutron pairing should be weak.

The neutrino luminosity due to Cooper pairing of nucleons may exceed that due
to the modified Urca process in normal matter by a factor of up to $\sim 100$.
Thus, neutrino emission produced
by Cooper pairing will not affect the cooling of
massive neutron stars
(if the direct Urca process is allowed
in the inner core and superfluidity becomes weak there, unable
to suppress fast neutrino emission).

In principle, the observations could be explained in the same manner as in
Sect.\ \ref{psf} but assuming strong neutron superfluidity 
in the stellar core
(instead of
strong proton one) and weak proton superfluidity (instead of
weak triplet-state neutron pairing).
This has been shown (e.g., Yakovlev et al.\ 1999)
in simplified cooling simulations with $T_{\rm cp}$
and $T_{\rm cnt}$ constant throughout stellar cores.
However, this
scenario seems less likely in view of theoretical estimates of gaps.  The
simultaneous presence of strong neutron and proton superfluids in the core
would greatly reduce the stellar heat capacity and initiate a rapid cooling of
low-mass
stars at $t \ga 3 \times 10^4$ yrs, in sharp contradiction
with the observations of old and warm sources.  Finally, very strong neutron
or proton superfluidity ($T_{\rm c} \ga 10^{10}$ K) everywhere throughout
neutron star cores would suppress the direct Urca process and produce slow
cooling of neutron stars of any mass
during the neutrino cooling stage (Yakovlev et al.\ 1999).
This too is in conflict with observation.

\subsection{Very Slowly Cooling Low-Mass Neutron Stars
and the Physics of the Crust}
\label{very slow cooling}

As discussed in Sect.\ \ref{psf},
low-mass stars with strong proton superfluidity
in their cores form a special class of {\it very slowly
cooling neutron stars}. Their cooling
is {\it insensitive to the physics of the core}:
to the equation of state,
the stellar mass, and to the model for proton superfluidity.
These stars differ from others by their very low
neutrino luminosity. As a result, their cooling
is {\it especially sensitive to the physics of the stellar crust},
since it is determined mainly by the presence of
({\it i}) singlet-state neutron superfluidity
in the inner crust, as well as by
the presence of ({\it ii})
accreted matter and ({\it iii}) magnetic fields
in the blanketing envelopes. All these effects are
of comparable strength. They are analyzed below
following Potekhin et al.\ (2003) with regard to
the observations of RX J0822--4300
and PSR B1055--52.

\begin{figure}[!t]        
\begin{center}
\leavevmode \epsfxsize=9.0cm
 \epsfbox[60 185 330 415]{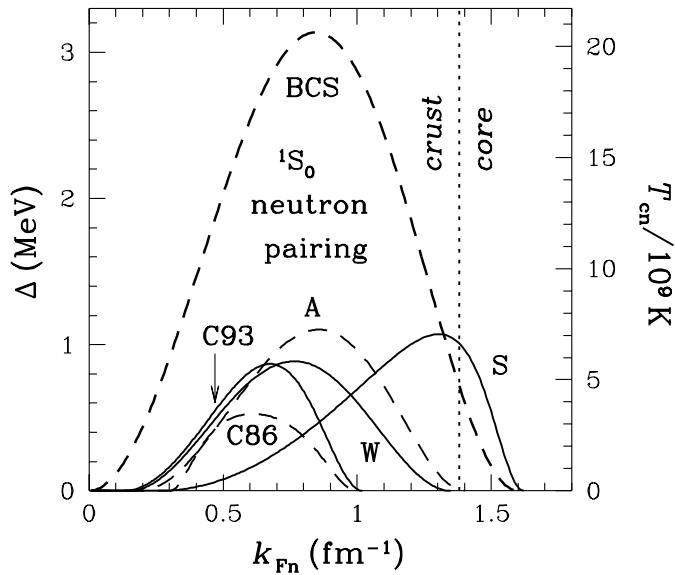}
\end{center}
\caption{
Energy gaps (left vertical axis) and
critical temperatures (right vertical axis)
for various models of crustal neutron pairing
(see text) as a function of
 neutron Fermi wave number. The vertical dotted
line marks the crust-core interface.
\label{fig:gap}}
\end{figure}

\begin{figure}[!t]        
\begin{center}
\leavevmode \epsfxsize=\textwidth
 \epsfbox[50 190 590 410]{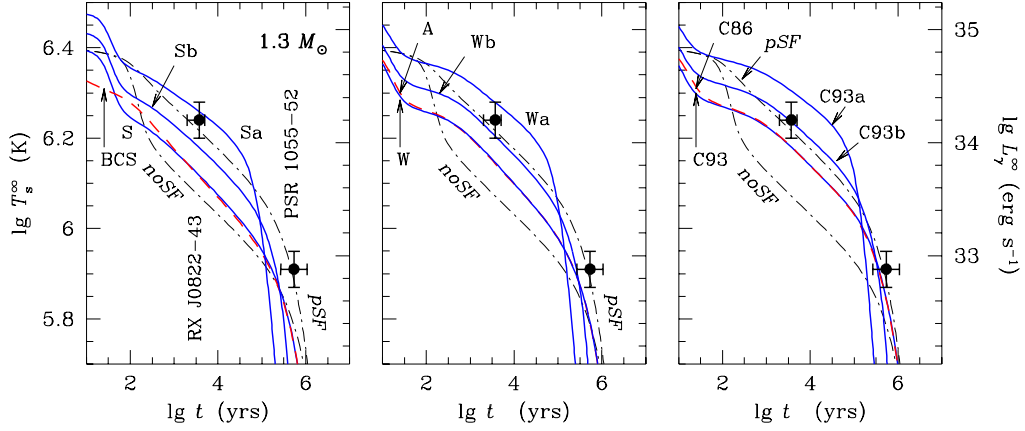}
\end{center}
\caption{
Evolution of redshifted effective surface
temperature $T^\infty_{\rm s}$ (left-hand scale) and
photon luminosity $L_\gamma^\infty$ (right-hand scale)
of a low-mass ($1.3\,M_\odot$)
neutron star
confronted with
observations of RX J0822--4300 and PSR B1055--52. Dot-dashed
curves: model without any superfluidity ({\it noSF}) or
with only strong proton superfluidity (model 1p) in the core ({\it
pSF}). Other curves
include the effects of proton superfluidity in the core
and a model of crustal neutron pairing from Fig.\ \ref{fig:gap}
(BCS or S: {\it left panel};
W or A: {\it center panel}; C86 or C93: {\it right panel}). Models
Sa, Wa, and C93a include also the effects of
fully accreted envelopes ($\Delta M/M=10^{-7}$).
Models Sb, Wb, and C93b
have nonaccreted envelopes and a dipolar magnetic
field $B_{\rm p}=10^{15}$ G.
\label{fig:lowm}}
\end{figure}

Our analysis is illustrated in Figs.\ \ref{fig:gap} and \ref{fig:lowm}.  For
definiteness, we take a 1.3 \msun\ neutron star with proton superfluidity
(model 1p) in the core and neglect
triplet-state neutron pairing in the
core (as discussed in Sect.\ \ref{pnsf}).  The two dot-dashed curves labeled
{\it noSF} and {\it pSF} in each panel of Fig.\ \ref{fig:lowm} show the
cooling of a nonsuperfluid star and
a one with strong proton superfluidity, both
of them without magnetic fields and accreted envelopes.  As shown in Sect.\
\ref{psf}, proton superfluidity delays the cooling, making it consistent with
the observations of RX J0822--4300 and PSR B1055--52.

The curves demonstrate how neutron
superfluidity in the crust initiates neutrino
emission due to the Cooper pairing of neutrons and noticeably
accelerates the cooling of low-mass neutron stars (Yakovlev et al.\ 2001b).

For example,
Fig.\ \ref{fig:gap} shows the dependence of the superfluid
gap $\Delta$ (left vertical axis) and the
associated critical temperature $T_{\rm cn}$
(in units of $10^9$ K, right axis)
on neutron Fermi wave number $k_{\rm Fn}$ (as a measure of density)
for six models of crustal superfluidity (from Lombardo \& Schulze 2001).
Model BCS
is the simplest model, in which the pairing interaction it taken to be
the neutron-neutron interaction in free space.
The five other models --
C86 
(Chen et al.\ 1986),
C93 
(Chen et al.\ 1993),
A   
(Ainsworth et al.\ 1989),
W   
(Wambach et al.\ 1993),
and
S   
(Schulze et al.\ 1996) --
include medium polarization effects which weaken the
pairing.  While all curves exhibit the same
qualitative
behavior, there are quantitatively important differences.
Model BCS is oversimplified, since it does not take into account effects of
the medium. In a medium, the gap is affected mainly by exchange of spin
fluctuations, which reduce the gap, just as they do in metals.  Model S
includes the effects of spin fluctuations, and the reason for it giving
results so different from the others is unclear.  Despite the superficial
similarity between the results of the other models, there is a spread of a
factor two in the predictions for the upper density at which neutron
superfluidity disappears.

Six cooling curves (BCS, S, A, W, C86, and C93)
in Fig.\ \ref{fig:lowm} are calculated adopting
the 1p model for proton superfluidity
in the core and one of the models
for
neutron superfluidity
in the crust.
Crustal superfluidity
accelerates the cooling and complicates
the interpretation of the observations of
RX J0822--4300 and PSR B1055--52.
The six curves are naturally divided into
three pairs shown in three panels of Fig.\ \ref{fig:lowm}.
The curves within each pair are very close,
while the pairs differ from one another.
The superfluid gaps
for any pair are different but disappear at the same density
(Fig.\ \ref{fig:gap}). This density restricts the volume in which
neutrino emission due to the singlet-state neutron pairing
may operate and accelerate the cooling.
For models BCS and S, superfluidity
penetrates into the stellar core,  for models
A and W it dies out at the crust-core interface,
while for models C86 and C93 it dies out
well before the interface.
Naturally,
 the effects of superfluidity are weakest for models C86 and
C93, while
models BCS and S, which are less realistic from a microscopic point of view,
produce
the most dramatic effect: the cooling curves lie much lower than the {\it pSF}
curve, and are marginally inconsistent with the data.

All neutron stars are expected to have the same superfluid properties
but may have different magnetic fields
and
envelopes. Figure \ref{fig:lowm} shows several
cooling curves, where the presence of magnetic fields
($B_{\rm p}=10^{15}$ G)
or accreted envelopes ($\Delta M/M=10^{-7}$)
is taken into account in addition to the
crustal superfluidity.
As discussed in Sect.\ \ref{magnetic fields},
magnetic fields and accreted envelopes
have opposite effects on $T_{\rm s}^\infty$ during the neutrino
and photon cooling stages. PSR B1055--52
is just passing from one cooling stage to the other
and has no
superstrong magnetic field. It is not expected to possess an
extended accreted envelope. Thus, the effects of magnetic fields
and accreted envelopes on the evolution of this pulsar are
thought to be minor, but they may be important for
RX J0822--4300. Including
these
effects, one can substantially raise the cooling curve, thereby bringing
it into agreement with observation.

These results are preliminary.  First, the observational data are not too
certain (Sect.\ \ref{observ}).  Second, the magnetic fields and accreted
envelopes have been considered as fixed.  In fact, the magnetic field strength
and geometry may evolve (particularly, due to ohmic decay) and the composition
of surface layers may change (e.g., due to diffusive nuclear burning, Chang \&
Bildsten 2003).  An important topic for future research is the self-consistent
simulation of the magnetic, chemical, and thermal evolution of neutron stars.

As seen from Fig.\ \ref{fig:lowm}, one can easily
build a model of a middle-aged ($t \la 10^4$ yr)
low-mass cooling star which would be noticeably hotter than
RX J0822--4300. It is sufficient to assume
strong proton superfluidity in the core,
crustal superfluidity according to model C86 or C93,
and a massive accreted envelope. This will give
a cooling curve similar to curve C93a in the
right
panel
of Fig.\ \ref{fig:lowm}. One can additionally
push this cooling curve up by assuming a very strong
magnetic field, $B_{\rm p} \ga 10^{15}$ G, but this
rise will be small. Thus, the C93a curve
is close to the {\it limiting cooling curve} for
the {\it hottest} neutron star of age $t \la 10^4$ yrs.
Notice that such a star will cool very quickly
during the photon cooling stage.

\subsection{Cooling of Old
Neutron Stars. Reheating Mechanisms}
\label{old neutron stars}

Let us outline the cooling of old neutron stars ($t \ga 10^5-10^6$ yrs,
the photon cooling stage). The problem is complicated.
Its important
ingredient is the $T_{\rm s}-T_{\rm b}$ relation.
In old stars the heat-insulating surface layer becomes
extremely thin.
The internal temperature is expected to
become close to the surface one (very roughly) at
$T_{\rm s} \la 10^3$ K. Such a cold surface may be solid;
its thermal emission may be reduced by the limited transparency
of surface material.

For a nonsuperfluid star without a magnetic field and without an
accreted envelope,
a temperature $T_{\rm s} \sim 10^3$ K
is reached at $t \sim 2 \times 10^7$ yrs independently
of stellar mass and the presence of the enhanced neutrino emission.
By that time the neutrino emission properties
(crucial for $t \la 10^5$ yrs) will be unimportant, and the
slow and rapid cooling curves will converge to a single curve.
Assuming black-body emission
with $T_{\rm s} \approx T_{\rm b}$
during the later cooling stages, from Eq.\ (\ref{therm-isotherm})
one finds the approximate cooling law $t
\sim
10~{\rm yr}/T_{\rm s6}^2$,
which gives $T_{\rm s} \sim 100$ K at $t=10^9$ yrs.
The cooling of old stars is affected by
superfluidity in the stellar core: strong superfluidity
of nucleons suppresses the heat capacity and
accelerates the cooling. For instance, strong superfluidity
of neutrons and protons reduces $T_{\rm s}(t)$ at
$t \ga 10^7$ yrs by a factor of $\sim$4.

These cooling scenarios
are idealized. Old neutron stars have a low heat capacity and therefore
any
weak reheating may drastically raise their temperatures.
One such mechanism for  reheating is {\it viscous dissipation of
rotational energy} within the star. Studies of this effect were initiated by
Alpar et al.\ (1987)
and
Shibazaki \& Lamb (1989) who
took into account
viscous dissipation due to the interaction
of superfluid and normal components of matter
in the inner crust.
The cooling theory with viscous reheating has been developed
further in a number of articles cited by Page (1998a, b)
and Yakovlev et al.\ (1999).

Reheating may also be produced by the
energy release due to {\it weak deviations
from beta equilibrium} in a neutron star core
(Reisenegger 1995).

Another possibility is that a star may be heated by {\it ohmic dissipation
of the magnetic field} in a nonsuperfluid core
due to the enhancement of the electrical resistivity
across a strong magnetic field. This mechanism was
suggested by Haensel et al.\ (1990), and
other references may be found in Yakovlev et al.\ (1999).
{\it Ohmic decay
of the magnetic field in the crust} can also heat the star
(Miralles et al.\ 1998, Urpin \& Konenkov 1998).

Reheating of old isolated neutron stars may also be
provided by accretion from the interstellar
medium or by pulsar activity.

As a rule, reheating mechanisms are
model dependent and can produce noticeable effects
only in old stars. Unfortunately, no reliable observational
data on the thermal states of such stars are yet available.
No reheating is required to interpret
the observations of middle-aged neutron stars (Sect.\ \ref{observ}).

\section{COOLING OF NEUTRON STARS WITH EXOTIC CORES}
\label{exotica}

\begin{figure}[!t]        
\begin{center}
\leavevmode \epsfxsize=13cm
 \epsfbox[20 155 450 415]{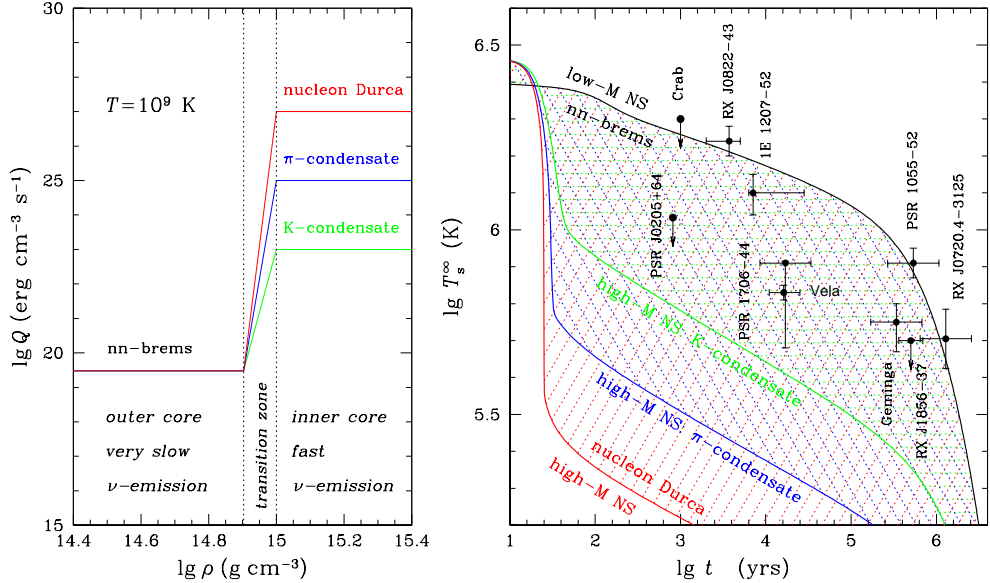}
\end{center}
\caption{{\it Left:} Schematic model for the density dependence
of the neutrino emissivity in a neutron star core
at $T=10^9$ K assuming very slow neutrino emission
in the outer core and three scenarios for fast
emission in the inner core. {\it Right:} Ranges of
$T_{\rm s}^\infty$ (single, double and triple
hatching) for the three types of fast emission,
compared with observations. Each range is limited
by the upper cooling curve which is for a low-mass star
and
a lower curve, which is
for a high-mass star. (See text.)}
\label{fig:exot}
\end{figure}

As the next step, following Yakovlev \& Haensel (2003),
we explore the hypothesis of
exotic matter in the cores
of neutron stars.
We adopt the model of
neutrino emission given by Eq.\ (\ref{Qnu}) and
shown in the left panel of Fig.\ \ref{fig:exot}.
Quite generally, we assume the presence of
an outer core with slow neutrino emission,
an inner core with fast neutrino emission,
and an intermediate zone ($\rho_{\rm s} \la \rho \la \rho_{\rm f}$).
Using this model we
obtain three types of cooling neutron stars similar to those
discussed (Sect.\ \ref{psf}) for stars with nucleon cores:
low-mass stars ($\rho_{\rm c} \la \rho_{\rm s}$)
which cool slowly,
high-mass stars ($\rho_{\rm c} \ga \rho_{\rm f}$),
which cool rapidly via enhanced neutrino emission
from the inner core, and medium-mass stars
($\rho_{\rm s} \la \rho_{\rm c} \la \rho_{\rm f}$),
whose cooling is intermediate.

The right panel of Fig.\ \ref{fig:exot} displays cooling curves for models of
low-mass and high-mass stars (no magnetic fields and accreted envelopes) for
three qualitatively different equations of state and compositions for the
cores, which lead to vastly different neutrino emission rates in the inner
cores.  We stress that the actual composition and equation of state of matter
should be the same for all neutron stars, provided there is sufficient time
for the matter to come into equilibrium, but at present they are unknown.

The upper cooling curve is for a low-mass star.
Such neutron stars are thought to consist of
nucleon matter; the cooling curves for all three equations of state
have to be nearly the same. For definiteness, we present the
cooling curve for a 1.35 \msun\ star with strongly superfluid
protons from the left panel of Fig.\ \ref{fig:1p2p}.  In this case, the main
neutrino emission is produced by neutron-neutron bremsstrahlung
in the core.

The three lower cooling curves in the right
panel of Fig.\ \ref{fig:exot}
refer to high-mass neutron stars with
different equations of state. The lowest curve corresponds
to
nucleon matter. For example, we present the
curve from Fig.\ \ref{fig:1p2p} for the maximum-mass
neutron star.  A similar curve should describe the cooling of
massive neutron stars with hyperonic cores. Two higher curves
in Fig.\ \ref{fig:exot}
schematically
show the
cooling of
massive neutron stars with a pion or kaon condensate
in the core (with $Q_{\rm f}=10^{25}$ or
$10^{23}$ erg cm$^{-3}$ s$^{-1}$ in Eq.\ (\ref{Qnu}),
respectively). Neutron stars with quark cores are expected to
show nearly the same cooling behavior as stars with kaon condensates.

For a given equation of state for dense matter,
the highest cooling curve corresponds to a low-mass star and
the lowest cooling curve to a high-mass star.  Between these two curves lies
a sequence of cooling curves for neutron stars with masses
between the highest and
lowest ones (hatched regions).
The data are consistent with any of the models for neutrino emission
(nucleon direct Urca, pion condensate or kaon condensate),
as many
authors have concluded, e.g., Page, 1998a, b). Obviously,
the discovery of cooler neutron stars would have important implications for
the composition of matter at supernuclear densities.

The rather uniform scatter of the observational
points suggests the existence of a
class of intermediate mass neutron stars. Their mass range is sensitive
(Yakovlev \& Haensel 2003)
to the position and width of the transition layer
(Fig.\ \ref{fig:exot})
between the slow and fast neutrino emission zones.
Unfortunately, these parameters cannot be
constrained
by the current data.
For instance, $\rho_{\rm s}$
can be placed anywhere between $\sim 8 \times 10^{14}$ to
$\sim 1.2 \times 10^{15}$ g cm$^{-3}$ for a broad range
of equations of state. The deduced masses of
medium-mass
stars will be different, but it will still be possible
to explain all the sources. For kaon-condensed
matter, the difference in rates between slow and fast neutrino emission
processes is not too large and, consequently, stars with a significant range
of stellar masses exhibit cooling behavior intermediate between the two
limiting cases
even if the transition
zone is absent ($\rho_{\rm f}=
\rho_{\rm s}$). In other cases the transition zone must
be rather wide ($\rho_{\rm f}-\rho_{\rm s} \ga 0.1 \rho_{\rm s}$)
to explain the medium-mass sources. In the scenarios
described in Sect.\ \ref{psf} the nonzero width of the
transition zone has been produced by the weakening
of proton superfluidity at high $\rho$.

Our analysis is a restricted one, but it has pointed to a number of
general features.  We have not described
in detail
the
cooling of neutron stars with quark cores, which is very rich
in physics; a comprehensive study has been carried
out by Schaab et al.\ (1996). Nor have
we considered the cooling of
bare strange stars. The physics of emission from the surface of these objects
is very different from that of ordinary neutron stars because
of the very high plasma frequency of surface quark layers.
The issue has recently been addressed by Page \& Usov (2002).

\section{THERMAL STATES OF TRANSIENTLY ACCRETING NEUTRON STARS}
\label{transients}

Now we discuss thermal states of accreting
neutron stars in soft X-ray transients
(SXRTs). We shall follow mainly the considerations
of Yakovlev et al.\ (2003, 2004b).
SXRTs undergo
periods of outburst activity
(lasting from days to months) superimposed on
quiescent periods (lasting from months to decades);
see, e.g., Chen et al.\ (1997).
Their activity is most probably regulated
by accretion from a disk around the neutron star.
During quiescence, when
accretion is absent or greatly suppressed,
some sources emit rather intense
thermal radiation, which indicates that the neutron stars are rather
hot. A possible explanation for these sources (Brown et al.\ 1998)
is that they are neutron stars powered by {\it deep crustal heating}
(Haensel \& Zdunik 1990, 2003) produced by nuclear transformations
in accreted matter as it sinks into
the inner crust 
under the weight of newly accreted material.
The total energy release is about 1.1--1.5 MeV per
accreted baryon,
and the total heating power
is $L_{\rm h}
\approx (6.6-9.0) \times 10^{33}  \,
\dot{M}/(10^{-10} \, \msun \; {\rm yr}^{-1})$
erg s$^{-1}$, where $\dot{M}$ is the mass accretion rate.
The main energy release is produced by pycnonuclear reactions
at $\rho \sim (10^{12}-10^{13})$
g cm$^{-3}$ (several hundred meters below the surface).
The heat is spread over the neutron star by thermal conduction
and radiated away by emission of photons from the surface and
neutrinos from the interior. Generally,
the {\it surface temperature depends on the internal
structure of the star}, and this gives a new method for studying the
internal structure.

\begin{figure}[!t]        
\begin{center}
\leavevmode \epsfxsize=9.5cm
 \epsfbox[80 215 550 680]{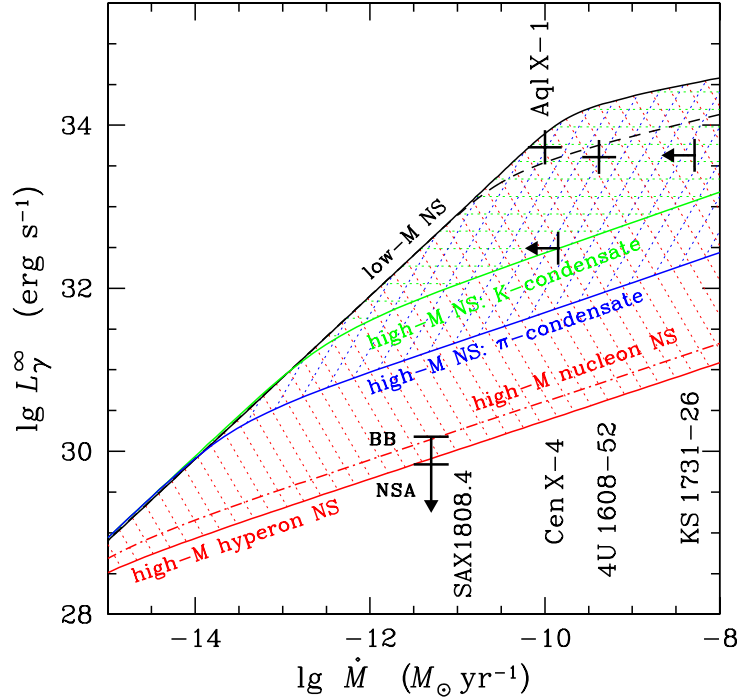}
\end{center}
\caption{Quiescent thermal luminosity of several
neutron stars in SXRTs versus mass accretion rate
compared with theoretical curves.
Three ranges of
$L_\gamma^\infty$ (single, double and triple
hatching) correspond to the three types of fast neutrino emission.
Each range is limited
by the upper heating curve of a low-mass star and
a lower curve of a high-mass star. See the text for details.}
\label{fig:trans}
\end{figure}

Neutron stars in SXRTs are
thermally inertial
objects with thermal relaxation times $\sim 10^4$ yr (Colpi et al.\ 2001).
Thus their internal temperatures $T_i$
(and hence their surface temperatures)
are insensitive to the transient nature of the accretion rate.
Accordingly, $T_i$ can
be determined by solving
the thermal balance equation (\ref{therm-isotherm})
in the steady-state approximation: $L_{\rm h}^\infty(\dot{M})=L_\nu^\infty
(T_i)+L_\gamma^\infty$,
where $\dot{M} \equiv \langle \dot{M} \rangle$
is the time-averaged accretion rate
and $L_\gamma^\infty$ is the quiescent thermal luminosity.
A solution gives
a {\it heating curve} --
a relationship between the thermal luminosity
$L_\gamma^\infty(\dot{M})$ as a function of accretion rate, or, equivalently,
$T_{\rm
s}^\infty(\dot{M})$. The heating curves of accreting neutron stars are closely
related to the cooling curves of isolated
neutron stars (e.g., Colpi et al.\ 2001, Yakovlev et al.\ 2003)
since they are determined by essentially the same physics.
The main difference is that the steady states
of accreting neutron stars are independent of the heat capacity of the star,
although the approach to such a state does depend
on the heat capacity.

By analogy with the cooling theory,
accreting neutron stars may operate in
the photon-emission regime ($L_\nu \ll L_\gamma \approx L_{\rm h}$)
or the neutrino-emission regime
($L_\nu \approx L_{\rm h}\gg L_\gamma$). The latter regime, which
is very sensitive to neutron star structure,
is realized at higher $\dot{M}$
(e.g., Yakovlev et al.\ 2003). For instance,
a low-mass nonsuperfluid neutron star with a core of neutrons, protons and
electrons enters
the neutrino-emission regime
for $\dot{M} \ga 3 \times 10^{-12}$ \msun\ yr$^{-1}$,
while a high-mass star does so for
$\dot{M} \ga 10^{-15}$ \msun\ yr$^{-1}$.

Just as in the case of cooling isolated neutron stars
(Sects.\ \ref{psf} and \ref{exotica}), there are three types of
accreting neutron stars in the neutrino-emission regime.
{\it Low-mass} stars have exceptionally low
neutrino emission; they are the hottest for the same $\dot{M}$.
{\it High-mass} stars have
much stronger neutrino emission and are the coldest ones.
{\it Medium-mass} stars are intermediate between the
hottest and the coldest ones.
The
$L_\gamma^\infty(\dot{M})$
curves are sensitive to the
presence in the surface layers
of accreted matter containing light elements
which remain
unburnt in thermonuclear reactions by the beginning of a quiescent stage.

Figure \ref{fig:trans} presents
the limiting
$L_\gamma^\infty(\dot{M})$
curves for three different
scenarios
of
neutrino emission, basically the
same as those studied in Sect.\ \ref{exotica}
(left panel of Fig.\ \ref{fig:exot}).
It is assumed that thermonuclear burning in
the surface layers produces Fe-like elements
beneath the unburnt matter.
The dashed curve is calculated
for the 1.1 \msun\
neutron star
whose core is composed of neutrons, protons and electrons
(the same equation of state as in Sect.\ \ref{nucleon cores}) with the strong
proton superfluidity model 1p. Actually, this curve is fairly insensitive
to the equation of state in the core, the neutron star mass (as long as
$\rho_{\rm c}
<
\rho_{\rm s}$), and the strong proton superfluidity model (just as for cooling
neutron stars, Sect.\ \ref{psf}). Thus, such curves obey the {\it
universality} rule, mentioned in Sect.\ \ref{nosf}. For accreting neutron
stars,
the
heating
curves are more universal when plotted in terms of
$L_\gamma^\infty(\dot{M})$ rather than $T_{\rm s}^\infty(\dot{M})$
(Yakovlev et al.\ 2004b).
The
uppermost solid
curve in Fig.\ \ref{fig:trans} is the same as the dashed
one but assumes the presence of $\Delta M =10^{-8}$ \msun\
of light elements on the surface. Light elements make the
plasma a better conductor of heat (Sect.\ \ref{TsTbrelation}) and thereby
increase $L_\gamma^\infty$ for a given $\dot{M}$.

The dot-dashed curve refers to the maximum-mass
(1.977 \msun) neutron star with the same equation of state and no accreted
envelope. This is the coolest accreting state for a given equation of state.
The even lower, solid curve refers to the maximum-mass
(1.975 \msun) neutron star with a hyperonic core (model 3
of Glendenning 1985 for the equation of state).
In this case
additional hyperonic
direct
Urca processes are
allowed.  They
increase the neutrino emission and make
the star
even colder,
closer to the  {\it limit of the coldest neutron stars with
nucleon-hyperon cores}. The singly hatched area in Fig.\ \ref{fig:trans}
is the region filled by heating curves
of neutron stars of various masses ($M \ga \msun$). It can be explained
by the models of transiently accreting neutron stars with nucleon-hyperon
cores. All the
$L_\gamma^\infty(\dot{M})$
curves described above are taken
from Yakovlev et al.\ (2004b).

The
second and third
lowest solid curves are
schematic models
(Yakovlev et al.\ 2003)
for high-mass neutron stars without accreted envelopes but
with pion-condensed or kaon-condensed cores, respectively
(with $Q_{\rm f}=10^{25}$ or $10^{23}$ erg cm$^{-3}$ s$^{-1}$,
as in Sect.\ \ref{exotica}).
The third curve is about
the same as for high-mass neutron stars with quark-matter cores.
Accordingly, the double and triple hatched regions
can be explained by models of accreting neutron stars
with pion-condensed and
kaon-condensed  (or quark-matter) cores.

These results are compared with observations of five SXRTs.
The data are the same as those taken by Yakovlev et al.\ (2003, 2004b).
We regard $L_\gamma^\infty$ as the thermal quiescent
luminosity of SXRTs, and take
the values of $L_\gamma^\infty$ for Aql X--1, Cen X--4,
4U 1608--552, KS 1731--26, and SAX 1808.4--3654
from Rutledge et al.\ (2002, 2001, 1999), Wijnands et al.\ (2002),
and Campana et al.\ (2002), respectively.
All these values (except that for SAX 1808.4--3654) have been
obtained with the aid of hydrogen atmosphere models.
The values of $\dot{M}$ for KS 1731--26 and Cen X--4
are most probably upper limits. Quiescent thermal
emission has not been detected
from SAX J1808.4--3658 and it is probably obscured
by a rather strong non-thermal emission.
We present the established upper limits of $L_\gamma^\infty$
obtained by Campana et al.\ (2002) from the observation
on March 24, 2001 (when the source was in a very low state),
using either the black-body (BB) or neutron star hydrogen atmosphere
(NSA) models.
Since all the data are rather uncertain we plot them
as large crosses or bars.

As seen from Fig.\ \ref{fig:trans}, we may interpret the neutron stars in 4U
1608--52 and Aql X--1 as low-mass stars with superfluid cores.  These stars
may be between the neutrino- and photon-emission regimes, while other neutron
stars are in the neutrino-emission regime.  The data on Aql X--1 are in better
agreement with the models of stars with light-element envelopes (see Yakovlev
et al.\ 2004b for details).  The neutron stars in Cen X--4 and SAX
J1808.4--3658 seem to require the fast neutrino cooling
and thus are
more massive.  The status of the neutron star in KS 1731--26 is less certain
because of the poorly determined $\dot{M}$; it too may require fast
neutrino emission.  Similar conclusions with respect to some of these sources
have been arrived at by a number of authors (cited in Yakovlev et al., 2003).

The observational point for Cen X--4
lies above (or near) all three limiting curves
for massive stars. Thus, we can consider the neutron star in Cen X--4 either
as a high-mass star (with a kaon-condensed or quark-matter core) or as
a medium-mass star (with a pion-condensed,
or nucleon-hyperon core). We shall be able
to explain all the data (except those for  the SAX source) on the basis of any
one of the three assumptions on the internal structure
(exactly as for cooling neutron stars in Sect.\ \ref{exotica}).

By contrast, the data
on SAX J1808.4--3658 indicate that the source contains
{\it a very cold neutron star}. Within the framework of our interpretation,
it can be explained {\it only as a high-mass neutron star with a
nucleon or nucleon-hyperon core} (and the nucleon-hyperon core is
preferable; see Yakovlev et al.\ 2004b, for
details).
However, this
conclusion is based on one observation of one source
and has to be confirmed in the future. Moreover,
the assumption that deep crustal burning of accreted matter
powers the quiescent thermal emission of SXRTs remains
a hypothesis. For instance, it seems that
a long-term variability of some X-ray transients
in quiescent states (e.g., Aql X--1
or MXB 1659--29, see Rutledge et al.\ 2002 or
Wijnands et al.\ 2004) cannot be associated with
deep crustal heating. Nevertheless, deep crustal
heating is a well established process (Haensel \& Zdunik 1990, 2003)
which is inevitable in accreting neutron stars and it must be taken
into account.

\section{CONCLUSIONS}

As a consequence of improved measurements of thermal emission from cooling
neutron stars in recent years, it has become very clear that the observations
cannot be explained on the basis of a single universal cooling curve.  If
thermal radiation from neutron stars in soft X-ray transient sources is due to
nuclear burning processes deep in the crust, the observations of isolated
neutron stars and X-ray transients can be analyzed within a common theoretical
framework.  Moreover, observations may be explained in terms of physically
reasonable models.

The basic ingredients of such a model are:

({\it a}) In the cores of massive neutron stars, a neutrino emission
process faster than the modified Urca one operates.
If one disregards the observations
of SAX J1808.4--3658, it is not possible to pin down which of the faster
processes
(direct Urca processes for nucleons and hyperons, a pion condensate, a kaon
condensate, or quark matter) is responsible, but if one includes the data from
SAX J1808.4--3658, the nucleon or hyperon direct Urca process would be
favored, and the other possibilities would be excluded.

({\it b}) In the cores of low-mass stars, neutrino emission
is slower than that produced by the modified Urca process. For instance,
this emission may be provided by
neutron-neutron bremsstrahlung while other potentially
efficient neutrino processes may be suppressed by strong superfluidity
of protons.

({\it c}) Medium-mass stars show cooling intermediate between slow
and fast.
In particular, they may cool via enhanced neutrino emission
partly suppressed by proton superfluidity.
The mass range for these stars is determined by the density range over
which the transition in the neutrino emission rate from slow to fast
occurs. Some physical models of neutron star interiors contradict
observations, for instance, the model of mild $^3$P$_2$ neutron
superfluidity in the stellar cores
with a maximum superfluid transition
temperature $T_{\rm cn}(\rho)$ in the range from
$2 \times 10^8$ to $2 \times 10^9$~K.

It is unlikely that advances in understanding the
nature of the interiors of neutron stars will come from a single piece of
evidence, but rather from a systematic appraisal of a variety of different
sorts of evidence, just as in many legal cases. Directions for future study
include:

$\bullet$ Further observations of thermal radiation from neutron stars.
A search for new very cold or very hot
stars would be useful. Very cold
neutron stars would rule out the possibility of
not too fast neutrino emission produced by
exotic matter in neutron star
cores.

$\bullet$ Further theoretical investigations of the effects of correlations
in dense matter.  In particular, the role of tensor correlations needs to
be reexamined following the work of Akmal \& Pandharipande (1997), which
found a strong increase of tensor correlations, a sign of incipient pion
condensation at relatively low densities,
and the recent study by Schwenk \&
Friman (2004) which pointed to the strong modification of the tensor force
by the nuclear medium.

$\bullet$ Information about neutron stars obtained from studies of cooling
needs to be integrated with what has been learned by other means.  Examples
are other observations of neutron stars,
for instance, measurements of their radii or gravitational redshifts.
Of special importance are observations of neutron stars 
in binary systems, which can be used to determine
neutron star masses.  Even a firm lower bound on a neutron star mass obtained
from, e.g., radio observations of compact binaries containing pulsars
(either binary neutron stars or
pulsar--white-dwarf binaries, such as
J0751+1807
reported recently by
Nice et al.\ 2004), could
rule out a number of theoretical equations of state.  It is also important
to ensure that the physical input to neutron star calculations is
consistent with experimental nuclear physics data on correlations between
nucleons, hyperons and other degrees of freedom in dense matter.

{\bf Acknowledgment.} We are grateful to
O.\ Gnedin, P.\ Haensel, A.D.\ Kaminker, K.\ Levenfish, A.\ Potekhin,
and A.\ Shibanov, DY's coauthors on papers discussed in this
review, and to G.G.\ Pavlov for enlightening remarks on
observational data. We are also grateful
to Olga Burstein and
Daniel Cordier for critical comments which improved the presentation
of the text and Table \ref{tab:param}.
This work has been supported partly by
the RFBR, grants 02-02-17668 and 03-07-90200.

\small
\noindent
Ainsworth~TL,  Wambach~J, Pines~D. 1989.
Effective interactions and superfluid energy gaps for low density neutron
matter.
{\it Phys.\ Lett.\ } B222:173--78

\noindent
Akmal A, Pandharipande VR. 1997. 
Spin-isospin structure and pion condensation in nucleon matter.
{\it Phys.\ Rev.\ } C56:2261--79

\noindent
Alford M. 2004.
Dense quark matter in Nature.
In {\it Finite Density QCD at Nara}. In press
[nucl-th/0312007]

\noindent
Alford M, Rajagopal K, Wilczek F. 1998.
QCD at finite baryon density: nucleon droplets
and color superconductivity.
{\it Phys.\ Lett.\ } B422:247--56

\noindent
Alpar A, Brinkmann W, \"{O}gelman H, Kizilo\u{g}lu \"{U},
Pines D. 1987.
A search for X-ray emission from a nearby pulsar -- PSR 1929+10.
{\it Astron.\ Astrophys.\ } 177:101--04


\noindent
Arendt RG, Dwek E, Petre R. 1991.
An infrared analysis of Puppis A.
{\it Astrophys.\ J.\ } 368:474--85

\noindent
Baiko DA,  Yakovlev DG. 1999.
Direct Urca process in strong magnetic fields and
neutron star cooling.
{\it Astron.\ Astrophys.\ } 342:192--200

\noindent
Bailin D,  Love A. 1984.
Superfluidity and superconductivity in relativistic fermion systems.
{\it Phys.\ Rep.\ }  107:325--85

\noindent
Balberg S,  Barnea N. 1998.
S-wave pairing of Lambda hyperons in dense matter.
{\it Phys.\ Rev.\ }  C57:409--16

\noindent
Baym G, Campbell DK. 1978. 
Chiral symmetry and pion condensation. In {\it Mesons in 
Nuclei}, ed. M Rho, D Wilkinson, Vol.\ III, pp.\ 1031--1094.
Amsterdam: North-Holland 

\noindent
Baym G, Pethick C, Pines D. 
1969. Superfluidity in neutron stars.
{\it Nature} 224:673--74

\noindent
Braje TM, Romani RW. 2002.
RX J1856-3754: Evidence for a stiff equation of state.
{\it Astrophys.\ J.\ } 580:1043--47

\noindent
Brisken WF, Thorsett SE, Golden A, Goss WM. 2003.
The distance and radius of the neutron star PSR B0656+14.
{\it Astrophys.\ J.\ }  593:L89--92

\noindent
Brown GE. 1995. 
Kaon condensation in dense 
matter. In: 
{\it Bose--Einstein Condensation}, ed.\ A~Griffin, DW 
Snoke, S Stringari, pp.\ 438--51. Cambridge: Cambridge Univ.\ 
Press

\noindent
Brown EF, Bildsten L, Rutledge RE. 1998.
Crustal heating and quiescent emission from transiently accreting neutron
stars.
{\it Astrophys.\ J.\ } 504:L95--98

\noindent
Burwitz V, Haberl F, Neuh\"auser R, Predehl P,
Tr\"umper J,  Zavlin VE. 2003.
The thermal radiation of the isolated neutron star
RX J1856.5-3754 observed with Chandra
and XMM-Newton.
{\it Astron.\ Astrophys.\ } 399:1109--14

\noindent
Campana S, Stella L, Gastaldello F, Mereghetti S,
Colpi M, Israel GL, Burderi L, Di Salvo T, 
Robba RN. 2002.
An XMM-Newton study of the 401 Hz accreting pulsar SAX J1808.4--3658 in
quiescence.
{\it Astrophys.\ J.\ } 575:L15--19

\noindent
Caraveo PA, Bignami GF, Mignani R, Taff LG. 1996.
Parallax observations with the Hubble Space Telescope
yield the distance to Geminga.
{\it Astrophys.\ J.\ }  461:L91--94

\noindent
Caraveo PA, De Luca A,
Mignani RP, Bignami GF. 2001.
The distance to the Vela pulsar gauged with Hubble Space Telescope
parallax observations.
{\it Astrophys.\ J.\ }  561:930--37

\noindent
Carter GW, Prakash M. 2002.
The quenching of the axial coupling in nuclear and neutron-star
matter.
{\it Phys.\ Lett.\ } B525:249--54

\noindent
Chang P, Bildsten L. 2003.
Diffusive nuclear burning in neutron star envelopes.
{\it Astrophys.\ J.\ } 585:464--74

\noindent
Chen~JMC, Clark JW, Krotscheck~E, Smith~RA. 1986.
Nucleonic superfluidity in neutron stars: $^1$S$_0$
neutron pairing in the inner crust.
{\it Nucl.\ Phys.\ } A451:509--40

\noindent
Chen JMC, Clark JW, Dav\'e RD, Khodel VV. 1993.
Pairing gaps in nucleonic superfluids.
{\it Nucl.\ Phys.\ } A555:59--89

\noindent
Chen W, Shrader CR, Livio M. 1997.
The properties of X-ray and optical light curves of X-ray novae.
{\it Astrophys.\ J.\ } 491:312--38

\noindent
Colpi M, Geppert U, Page D,  Possenti A. 2001.
Charting the temperature of the hot neutron star in a soft X-ray transient.
{\it Astrophys.\ J.\ } 548:L175--78

\noindent
Dodson D, Legge D, Reynolds JE, McCulloch PM. 2003.
The Vela Pulsar's proper motion and parallax derived from VLBI observations.
{\it Astrophys.\ J.\ } 596:1137--41

\noindent
Flowers EG, Ruderman M,  Sutherland PG. 1976.
Neutrino pair emission from finite-temperature
neutron superfluid and the cooling of young neutron stars.
{\it Astrophys.\ J.\ }  205:541--44

\noindent
Gamow G, Schoenberg M. 1941.
Neutrino theory of stellar collapse.
{\it Phys.\ Rev.\ } 59:539--47

\noindent
Gnedin~OY, Yakovlev DG, Potekhin AY. 2001.
Thermal relaxation in young neutron stars.
{\it Mon.\ Not.\ Roy.\ Astron.\ Soc.\ } 324:725--36

\noindent
Glen G,  Sutherland P. 1980.
On the cooling of neutron stars.
{\it Astrophys.\ J.\ } 239:671--84

\noindent
Glendenning NK. 1985.
Neutron stars are giant hypernuclei?
{\it Astrophys.\ J.\ } 293:470--93

\noindent
Glendenning N. 1996.
Compact Stars. Nuclear Physics, Particle Physics and
General Relativity.
Springer-Verlag: New York

\noindent
Gudmundsson EH, Pethick CJ, Epstein RI. 1983.
Structure of neutron star envelopes.
{\it Astrophys.\ J.\ } 272:286--300

\noindent
Gusakov ME. 2002.
Neutrino emission from superfluid
neutron-star cores: various types of neutron pairing.
{\it Astron.\ Astrophys.\ } 389:702--15

\noindent
Haensel P. 2003.
Equation of state of dense matter and maximum mass of
neutron stars.
In {\it Final Stages of Stellar Evolution},
ed.\ C Motch, J-M Hameury, pp.\ 249--84.
EAS Publications Series: EDP
Sciences. 

\noindent
Haensel P,  Zdunik JL. 1990.
Non-equilibrium processes in the crust of an accreting neutron star.
{\it Astron.\ Astrophys.\ } 227:431--36

\noindent
Haensel P, Zdunik JL. 2003.
Nuclear composition and heating in accreting neutron-star crusts.
{\it Astron.\ Astrophys.\ } 404:L33--36

\noindent
Haensel P. Urpin VA, Yakovlev~DG. 1990.
Ohmic decay of internal magnetic fields in neutron stars.
{\it Astron.\ Astrophys.\ }  229:133--37

\noindent
Halpern JP, Wang FY-H. 1997.
A broadband X-ray study of the Geminga pulsar.
{\it Astrophys.\ J.\ } 477:905--15

\noindent
Hanhart C, Phillips DR, Reddy S. 2001.
Neutrino and axion emissivities of neutron stars
from nucleon-nucleon scattering data.
{\it Phys.\ Lett.\ } B499:9--15

\noindent
Hoffberg M, Glassgold AE, Richardson RW,
Ruderman M. 1970.
Anisotropic superfluidity in neutron star matter.
{\it Phys.\ Rev.\ Lett.\ } 24:775--77

\noindent
Friman BL, Maxwell OV. 1979.
Neutrino emissivities of neutron stars.
{\it Astrophys.\ J.\ } 232:541--57

\noindent
Jaikumar P, Prakash M. 2001.
Neutrino pair emission from Cooper pair breaking
and recombination in superfluid quark matter.
{\it Phys.\ Lett.\ } B516:345--52

\noindent
Kaminker AD, Haensel P,  Yakovlev DG. 2001.
Nucleon superfluidity vs.\ observations of cooling
neutron stars.
{\it Astron.\ Astrophys.\ } 373:L17--20

\noindent
Kaminker AD, Yakovlev DG, Gnedin OY. 2002.
Three types of cooling superfluid neutron stars:
Theory and observations.
{\it Astron.\ Astrophys.\ } 383:1076--87

\noindent
Kaplan DB, Nelson AE. 1986.
Strange goings on in dense nucleonic matter.
{\it Phys.\ Lett.\ }  B175:57--63.
Erratum: 179:409

\noindent
Kaplan DL, Kulkarni SR, van Kerkwijk MH, \
Marshall HL. 2002.
X-ray timing of the enigmatic neutron star RX J0720.4-3125.
{\it Astrophys.\ J.\ }  570:L79--L83

\noindent
Lattimer JM, Prakash M. 2001.
Neutron star structure and the equation of state.
{\it Astrophys.\ J.\ } 550:426--42

\noindent
Lattimer JM, Pethick CJ, Prakash M, 
Haensel P. 1991.
Direct Urca process in neutron stars.
{\it Phys.\ Rev.\ Lett.\ }  66:2701--04

\noindent
Lattimer JM, van Riper KA, Prakash M, Prakash M.
1994. Rapid cooling and the structure of neutron stars.
{\it Astrophys.\ J.\ } 425:802--13

\noindent
Lombardo U, Schulze H-J. 2001.
Superfluidity in neutron star matter.
In {\it Physics of Neutron Star Interiors},
ed.\ D Blaschke, NK Glendenning, A Sedrakian,
pp.\ 30--53. Springer: Berlin

\noindent
Lorenz CP, Ravenhall DG, Pethick CJ. 1993.
Neutron star crusts.
{\it Phys.\ Rev.\ Lett.\ }  70:379--82

\noindent
Lyne AG, Pritchard RS, Graham-Smith F,  Camilo F. 1996.
Very low braking index for the
Vela pulsar.
{\it Nature} 381:497--98

\noindent
McGowan KE, Zane S, Cropper M, Kennea JA, C\'ordova FA,
Ho C, Sasseen T,  Vestrand WT. 2004.
XMM-Newton observations of PSR B1706--44.
{\it Astrophys.\ J.\ } 600:343--50

\noindent
Migdal AB. 1959.
Superfluidity and the moments of inertia of nuclei.
{\it Nucl.\ Phys.\ } 13:655--74

\noindent
Migdal AB. 1971.
Stability of vacuum and limiting fields.
{\it Uspekhi Fiz.\ Nauk} 105:781--81

\noindent
Miralles~JA, Urpin~VA, Konenkov~DYu. 1998.
Joule heating and the thermal evolution of old neutron stars.
{\it Astrophys.\ J.\ } 503:368--73

\noindent
Motch C, Zavlin VE, Haberl F. 2003.
The proper motion and energy distribution of the isolated neutron star.
{\it Astron.\ Astrophys.\ } 408:323--30

\noindent
Negele JW, Vautherin D. 1973.
Neutron star matter at sub-nuclear densities.
{\it Nucl.\ Phys.\ }  A207:298--320

\noindent
Nice DJ, Splaver EM, Stairs IH. 2004.
Heavy neutron stars? A status report on Arecibo timing
of four pulsar -- white dwarf systems.
In {\it Young Neutron Stars and Their Environments},
IAU Symposium No.\ 218, ed.\ F Camilo, BM Gaensler.
In press [astro-ph/0311296]

\noindent
Oyamatsu K. 1993.
Nuclear shapes in the inner crust of a neutron star.
{\it Nucl.\ Phys.\ } A561:431--52

\noindent
Page D. 1993.
The Geminga neutron star: evidence for nuclear superfluidity
at very high density.
In {\it Proc.\ of the First Symposium on Nuclear
Physics in the Universe}, ed.\
MR Strayer, MW Guidry, p.\ 15.
Bristol: Adam Hilger

\noindent
Page D. 1995.
Surface temperature of a magnetized neutron star
and interpretation of the ROSAT data. 1: Dipole fields.
{\it Astrophys.\ J.\ } 442:273--85

\noindent
Page D. 1998a.
Thermal evolution of isolated neutron stars.
In {\it The Many Faces of Neutron Stars},
ed.\ R Buccheri, J van Paradijs,  MA Alpar,
pp.\ 539--52. Kluwer: Dordrecht

\noindent
Page D. 1998b.
Thermal evolution of isolated neutron stars.
In {\it Neutron Stars and Pulsars},
ed.\ N Shibazaki, N Kawai, S Shibata, T Kifune,
pp.\ 183--94. Universal Academy Press: Tokyo

\noindent
Page D,  Applegate JH. 1992.
The cooling of neutron stars by the direct Urca process.
{\it Astrophys.\ J.\ } 394:L17--20

\noindent
Page D,  Usov VV. 2002.
Thermal evolution and light curves of young bare strange stars.
{\it Phys.\ Rev.\ Lett.\ } 89:131101

\noindent
Pavlov GG. 2003. Private communication

\noindent
Pavlov GG, Zavlin VE. 2003.
Thermal radiation from cooling neutron stars.
In {\it Texas in Tuscany. XXI Texas Symposium on Relativistic Astrophysics},
ed.\ R Bandiera, R Maiolino, F Mannucci, pp.\ 319--28. Singapore: 
World Scientific Publishing

\noindent
Pavlov GG, Zavlin VE, Sanwal D,
Burwitz V, Garmire GP. 2001.
The X-Ray spectrum of the Vela pulsar resolved with the Chandra X-ray
observatory.
{\it Astrophys.\ J.\ } 552:L129--33

\noindent
Pavlov GG, Zavlin VE, Sanwal D, Tr\"umper J. 2002a.
1E 1207.4--5209: The puzzling pulsar at the center of the
supernova remnant PKS 1209--51/52.
{\it Astrophys.\ J.\ } 569:L95--98

\noindent
Pavlov GG, Zavlin VE, Sanwal D. 2002b.
Thermal radiation from neutron stars: Chandra results.
In {\it 270 WE-Heraeus Seminar on
Neutron Stars, Pulsars and Supernova Remnants},
ed.\ W Becker, H Lesh, J Tr\"umper, pp.\ 273--86.
MPE: Garching 

\noindent
Pethick CJ. 1992.
Cooling of neutron stars.
{\it Rev.\ Mod.\ Phys.\ } 64:1133--40

\noindent
Pethick CJ,  Ravenhall DG. 1995.
Matter at large neutron excess and the physics of neutron-star crusts.
{\it Ann.\ Rev.\ Nucl.\ Particle Sci.\ } 45:429--84

\noindent
Pons JA, Miralles JA, Prakash M, Lattimer JM. 2001.
Evolution of proto-neutron stars with kaon condensates.
{\it Astrophys. J.\ } 553:382--93

\noindent
Potekhin AY, Yakovlev DG. 2001.
Thermal structure and cooling of neutron stars
with magnetized envelopes.
{\it Astron.\ Astrophys.\ } 374:213--26

\noindent
Potekhin~AY, Chabrier G, Yakovlev DG. 1997.
Internal temperatures and cooling of neutron stars
with accreted envelopes.
{\it Astron.\ Astrophys.\ } 323:415--28

\noindent
Potekhin AY, Yakovlev DG, Chabrier G, Gnedin OY. 2003.
Thermal structure and cooling of superfluid neutron stars with accreted
magnetized envelopes.
{\it Astrophys.\ J.\ } 594:404--18

\noindent
Prakash M, Ainsworth TL,  Lattimer JM. 1988.
Equation of state and the maximum mass of neutron stars.
{\it Phys.\ Rev.\ Lett.\ } 61:2518--21

\noindent
Prakash M, Prakash M, Lattimer JM, Pethick CJ. 1992.
Rapid cooling of neutron stars by hyperons and $\Delta$-isobars.
{\it Astrophys.\ J.\ }  390:L77--80

\noindent
Rutledge RE, Bildsten L, Brown EF, Pavlov GG, Zavlin VE.
1999.
The thermal X-Ray spectra of Centaurus X--4, Aquila X--1, and 4U 1608--522 in
quiescence.
{\it Astrophys.\ J.\ } 514:945--51

\noindent
Rutledge RE, Bildsten L, Brown EF, Pavlov GG, Zavlin VE.
2000.
A method for distinguishing between transiently accreting neutron stars and
black holes, in quiescence.
{\it Astrophys.\ J.\ }  529:985--96

\noindent
Rutledge RE, Bildsten L, Brown EF, Pavlov GG, Zavlin VE.
2002.
Variable thermal emission from Aquila X-1 in quiescence.
{\it Astrophys.\ J.\ } 577:346--58


\noindent
Reisenegger A. 1995.
Deviation from chemical equilibrium due to spin-down
as an internal heat source in neutron stars.
{\it Astrophys.\ J.\ }  442:749--57

\noindent
Sanwal D, Pavlov GG, Zavlin VE, Teter MA. 2002.
Discovery of absorption features in the X-ray spectrum
of an isolated neutron star.
{\it Astrophys.\ J.\ } 574:L61--64

\noindent
Sawyer RF. 1972.
Condensed $\pi^-$ phase of neutron star matter.
{\it Phys.\ Rev.\ Lett.\ } 29:382--85

\noindent
Scalapino DJ. 1972.
$\pi^-$ condensate in dense nuclear matter.
{\it Phys.\ Rev.\ Lett.\ } 29:386--88

\noindent
Schaab Ch, Weber F, Weigel MK, Glendenning NK. 1996.
Thermal evolution of compact stars.
{\it Nucl.\ Phys.\ } A605:531--65

\noindent
Schulze H-J, Cugnon J, Lejeune A, Baldo M, 
Lombardo U. 1996.
Medium polarization effects on neutron matter superfluidity.
{\it Phys.\ Lett.\ } B375:1--8

\noindent
Schwenk A., Friman B. 2004.
Polarization contributions to the spin-dependence of
the effective interaction in neutron matter.
{\it Phys.\ Rev.\ Lett.\ } in press
[nucl-th/0307089]

\noindent
Schwenk A, Jaikumar P, Gale C. 2004.
Neutrino bremsstrahlung in neutron matter from effective
nuclear interactions.
{\it Phys.\ Lett.\ B } in press
[nucl-th/0309072]

\noindent
Shapiro SL,  Teukolsky SA. 1983.
Black Holes, White Dwarfs, and Neutron Stars.
Wiley-Interscience: New York

\noindent
Shibazaki N,  Lamb FK. 1989.
Neutron star evolution with internal heating
{\it Astrophys.\ J.\ } 346:808--22

\noindent
Slane PO, Helfand DJ, Murray SS. 2002.
New constraints on neutron star cooling from
Chandra observations of 3C 58.
{\it Astrophys.\ J.\ } 571:L45--49

\noindent
Takatsuka T, Tamagaki R. 1995.
Nucleon superfluidity in kaon-condensed neutron stars.
{\it Progr.\ Theor.\ Phys.\ }  94:457--61

\noindent
Takatsuka T., Tamagaki R. 1997.
Effects of charged-pion condensation on neutron $^3$P$_2$ superfluidity.
{\it Progr.\ Theor.\ Phys.\ } 97:263--81

\noindent
Thorne KS. 1977.
The relativistic equations of stellar structure and evolution.
{\it Astrophys.\ J.\ } 212:825--31

\noindent
Tsuruta S. 1998.
Thermal properties and detectability of neutron stars. II.
Thermal evolution of rotation-powered neutron stars.
{\it Phys.\ Rep.\ } 292:1--130

\noindent
Tsuruta~S,  Cameron AGW. 1966.
Cooling and detectability of neutron stars.
{\it Canad.\ J.\ Phys.\ } 44:1863--94

\noindent
Urpin~VA, Konenkov~DYu. 1998.
Magnetic evolution of neutron stars.
In  {\it  Neutron Stars and Pulsars},
ed.\ N Shibazaki, N Kawai, S Shibata,  T Kifune,
pp.~171--78. Universal Academy Press: Tokyo

\noindent
Ushomirsky G, Rutledge RE. 2001.
Time-variable emission from transiently
accreting neutron stars in quiescence due to deep
crustal heating.
{\it Mon.\ Not.\ Roy.\ Astron.\ Soc.\ } 325:1157--66

\noindent
van Dalen ENE, Dieperink AEL,  Tjon JA. 2003.
Neutrino emission in neutron stars.
{\it Phys.\ Rev.\ } C67:065807

\noindent
Walter FM,  Lattimer JM. 2002.
A revised parallax and its implications for RX J185635--3754.
{\it Astrophys.\ J.\ } 576:L145--48

\noindent
Wambach~J, Ainsworth~TL, Pines~D. 1993.
Quasiparticle interactions in neutron matter for applications
in neutron stars.
{\it Nucl.\ Phys.\ } A555:128--50

\noindent
Weber F. 1999.
{\it Pulsars as Astrophysical Laboratories for Nuclear and
Particle Physics}.
Institute of Physics Publishing: Bristol

\noindent
Weisskopf MC, O'Dell SL,  Paerels F,
Becker W, Tennant AF,  Swartz D. 2004.
Chandra phase-resolved X-ray spectroscopy of the Crab pulsar.
{\it Astrophys.\ J.\ } In print [astro-ph/0310332]

\noindent
Wijnands R, Guainazzi M, van der Klis M, M\'endez M. 2002.
XMM-Newton observations of the neutron star X-ray transient KS 1731--260 in
quiescence.
{\it Astrophys.\ J.\ } 573:L45--49

\noindent
Wijnands R, Homan J, Miller JM, Lewin WHG. 2004.
Monitoring Chandra observations of the quasi-persistent neutron-star
X-ray transient MXB 1659--29 in quiescence: the cooling
curve of the heated neutron-star crust.
{\it Astrophys.\ J.\ } submitted [astro-ph/0310612]

\noindent
Winkler PF, Tuttle JH, Kirshner RP, Irwin MJ. 1988.
Kinematics of oxygen-rich filaments in Puppis A.
In {\it Supernova Remnants and the Interstellar Medium},
ed.\ RS~Roger, TL~Landecker, p.~65.
Cambridge: Cambridge Univ.\ Press

\noindent
Wolf RA. 1966.
Some effects of the strong interactions
on the properties of neutron-star matter.
{\it Astrophys.\ J.\ }  145:834--41

\noindent
Yakovlev DG, Haensel P. 2003.
What we can learn from observations of cooling neutron stars.
{\it Astron.\ Astrophys.\ } 407:259--64

\noindent
Yakovlev DG, Levenfish KP, Shibanov YuA. 1999.
Cooling of neutron stars and superfluidity in their cores.
{\it Uspekhi Fiz.\ Nauk} 169:825--68
(English translation: {\it Physics -- Uspekhi} 42:737--78)

\noindent
Yakovlev DG, Kaminker AD, Gnedin OY, Haensel P. 2001a.
Neutrino emission from neutron stars.
{\it Phys.\ Rep.\ }  354:1--155

\noindent
Yakovlev~DG, Kaminker AD, Gnedin OY. 2001b.
$^1$S$_0$ neutron pairing vs.\ observations of cooling
neutron stars.
{\it Astron.\ Astrophys.\ } 379:L5--8

\noindent
Yakovlev DG, Kaminker AD, Haensel P, Gnedin OY.
2002a, The cooling neutron star in 3C 58.
{\it Astron.\ Astrophys.\ } 389:L24--27

\noindent
Yakovlev DG, Gnedin OY, Kaminker AD, Potekhin AY. 2002b.
Cooling of superfluid neutron stars.
In {\it 270 WE-Heraeus Seminar on
Neutron Stars, Pulsars and Supernova Remnants},
ed.\ W Becker, H Lesh, J Tr\"umper, pp.\ 287--99.
MPE: Garching

\noindent
Yakovlev DG, Levenfish KP,  Haensel P. 2003.
Thermal state of transiently accreting neutron stars.
{\it Astron.\ Astrophys.\ } 407:265--71

\noindent
Yakovlev DG, Gnedin OY, Kaminker AD, Levenfish KP,
Potekhin AY. 2004a.
Neutron star cooling: theoretical aspects and observational
constraints.
{\it Adv.\ Space Res.\ } In press [astro-ph/0306143]

\noindent
Yakovlev DG, Levenfish KP, Potekhin AY, Gnedin OY,
Chabrier G. 2004b.
Thermal states of coldest and hottest neutron stars in
soft X-ray transients.
{\it Astron.\ \& Astrophys.\ } In press [astro-ph/0310259]

\noindent
Zavlin VE,  Pavlov GG. 2002.
Modeling neutron star atmospheres.
In {\it 270 WE-Heraeus Seminar on
Neutron Stars, Pulsars and Supernova Remnants},
ed.\ W Becker, H Lesh,  J Tr\"umper, pp.\ 263--72.
MPE: Garching

\noindent
Zavlin VE, Pavlov GG. 2004.
XMM observations of three middle-aged pulsars.
{\it Memorie della Societa' Astronomica Italiana.} In press
[astro-ph/0312326]

\noindent
Zavlin VE, Pavlov GG, Tr\"umper J. 1998.
The neutron star in the supernova remnant PKS 1209--52.
{\it Astron.\ Astrophys.\ } 331:821--28

\noindent
Zavlin VE, Tr\"umper J, Pavlov GG. 1999.
X-ray emission from the radio-quiet neutron star in Puppis A.
{\it Astrophys.\ J.\ } 525:959--67

\end{document}